\documentclass[11pt]{article}
\usepackage{latexsym,times}
\usepackage{amsmath}
\usepackage{amsxtra}
\usepackage{amscd}
\usepackage{amsthm}
\usepackage{amsfonts}
\usepackage{amssymb}
\usepackage{dsfont}
\usepackage{xcolor}
\usepackage{cancel}

\usepackage{framed}

\usepackage{esint}

\usepackage{enumitem}
\usepackage{float}

\usepackage{hyperref}
\usepackage[square,sort&compress,comma,numbers]{natbib}

\makeatletter

\@addtoreset{equation}{section} \makeatother

\input amssym.def
\input amssym.tex

\newcommand{\half}{{{\textstyle\frac{1}{2}}}}
\newcommand{\quarter}{{{\textstyle\frac{1}{4}}}}
\newcommand{\be}{\begin{equation}}
\newcommand{\ee}{\end{equation} }
\newcommand{\beqa}{\begin{eqnarray} }
\newcommand{\eeqa}{\end{eqnarray} }
\newcommand{\ba}{\begin{array}}
\newcommand{\ea}{\end{array}}
\newcommand{\bpm}{\begin{pmatrix}}
\newcommand{\epm}{\end{pmatrix}}

\newcommand{\SO}{\mathbf{SO}}

\newcommand{\ODD}{\mathbf{O}(D,D)}

\newcommand\rd{{\rm d}}

\newcommand\rdA{D}

\newcommand\cA{{\cal A}}

\newcommand\cD{{\cal D}}
\newcommand\cE{{\cal E}}

\newcommand\cH{{\cal H}}

\newcommand\cJ{{\cal J}}

\newcommand\cL{{\cal L}}

\newcommand\cO{{\cal O}}
\newcommand\cP{{\cal P}}

\newcommand\cS{{\cal S}}

\newcommand\brcP{{\bar{\cal{P}}}}

\newcommand\hcL{{\hat{\cal L}}}

\newcommand\dis{\displaystyle}

\newcommand\seceq{=}

\def\tx{\tilde{x}}

\def\tX{\tilde{X}}

\def\brDelta{{{\bar{\Delta}}}}

\def\brP{{\bar{P}}}

\newcommand{\DO}{\mathbf{\nabla}}

\newcommand{\na}{{\nabla}}

\newcommand{\fD}{\mathfrak{D}}
\newcommand{\brfD}{\bar{\fD}}

\newcommand\p\partial
\newcommand\req[1]{(\ref{#1})}

\begin{document}

\begin{titlepage}
\title{\vskip -100pt
\vskip 2cm 
Dynamics of Perturbations in Double Field Theory \\ \& \\Non-Relativistic String Theory \\}
\author{\sc{ Sung Moon Ko${}^{\dagger}$  ,~ Charles M. Melby-Thompson${}^{\sharp\ast}$ ,~Ren\'e Meyer${}^{\sharp}$ ~and~ Jeong-Hyuck Park${}^{\dagger}$}}
\date{}
\maketitle \vspace{-1.0cm}
\begin{center}
~\\
${}^{\dagger}$Department of Physics, Sogang University,  Seoul 121-742, Korea\\
${}^{\sharp}$Kavli Institute for the Physics and Mathematics of the Universe (WPI), The University of Tokyo Institutes for Advanced Study (UTIAS), The University of Tokyo, Kashiwanoha, Kashiwa, 277-8583, Japan\\
${}^{\ast}$Department of Physics, Fudan University, 220 Handan Road, 200433 Shanghai, China\\
~\\
\texttt{sinsmk2003@sogang.ac.kr\qquad  \quad charles.melby@ipmu.jp \\ rene.meyer@ipmu.jp\qquad                  \quad\quad\quad park@sogang.ac.kr  }
~~~\\
\end{center}
\begin{abstract}
\noindent    
Double Field Theory provides a geometric framework capable of describing string theory backgrounds that cannot be understood purely in terms of Riemannian geometry -- not only globally (`non-geometry'), but even locally (`non-Riemannian').
In this work, we show that the non-relativistic closed string theory of Gomis and Ooguri  \cite{Gomis:2000bd} arises precisely as such a non-Riemannian string background, 
and that the Gomis-Ooguri sigma model is equivalent to the Double Field Theory sigma model of \cite{Lee:2013hma} on this background.
We further show that the target-space formulation of Double Field Theory on this non-Riemannian background correctly reproduces the appropriate sector of the Gomis-Ooguri string spectrum. 
To do this, we develop a general semi-covariant formalism describing perturbations in Double Field Theory.  
We derive compact expressions for the linearized equations of motion around a generic on-shell background, and construct the corresponding fluctuation Lagrangian in terms of novel completely covariant second order differential operators. 
We also present a new non-Riemannian solution featuring Schr\"odinger conformal symmetry. 
\end{abstract} 
{\small
\begin{flushright}
~~\\
\textit{Preprint}: IPMU15-0123\\
\end{flushright}}
\thispagestyle{empty}
\end{titlepage}
\newpage
\tableofcontents 


\newpage

\section{Introduction}

While superstring theory in ten dimensions has only a few maximally symmetric vacua (ten dimensional flat Minkowski spacetime, pp-wave backgrounds, Anti de Sitter spacetime), its supersymmetric compactification down to four spacetime dimensions over a six-dimensional internal manifold already exhibits a vast number of string vacua \cite{Douglas:2003um,Ashok:2003gk}. 
All these vacua have in common that they are `geometric': they are described in terms of local coordinate patches glued by the transition functions of differential geometry, and equipped with a (pseudo-)Riemannian metric and other fields transforming in various representations of the Lorentz group.
%
Besides such geometric vacua, it is not unreasonable to expect that string theory as a theory of quantum gravity may also allow configurations that are intrinsically `non-geometric', \emph{i.e.,} ones that cannot be understood within the framework of Riemannian geometry. 
Indeed, large classes of what one could call `mildly non-geometric' compactifications down to four spacetime dimensions, \textit{i.e.,}~compactifications which are locally geometric but have global non-geometric features, have been identified \cite{Hellerman:2002ax,Shelton:2005cf}.
If we believe that string theory describes the universe we live in, we hence  cannot ignore the possibility that we live in a vacuum of string theory with genuinely non-geometric structure. The question of how many vacua or classes thereof, and of which nature, string theory actually allows, is hence of fundamental importance. In order to answer this question, we need formulations of string theory which go beyond usual supergravity and hence can describe non-geometric backgrounds. Double Field Theory (DFT) \cite{Siegel:1993xq,Siegel:1993th,Hull:2009mi,Hull:2009zb,Hohm:2010jy,Hohm:2010pp} is such a formulation.\footnote{For further guidance to the literature, we refer readers to Refs.\cite{Aldazabal:2013sca,Berman:2013eva,Hohm:2013bwa}.}  In this work we will use DFT to describe a particularly interesting kind of non-geometric background on which  the non-relativistic string theory \textit{\`a la} Gomis and Ooguri~\cite{Gomis:2000bd} is formulated in a systematic way. We will show \textit{i)}  how its CFT fits into a the DFT sigma model of~\cite{Lee:2013hma},  \textit{ii)}  how it arises as a solution of the DFT target space equations of motion,  and    that \textit{iii)} this target space theory correctly reproduces the string worldsheet spectrum of excitations described in \cite{Gomis:2000bd}. 
For earlier work on DFT or doubled sigma models, see refs.~\cite{Duff:1989tf,Tseytlin:1990nb,Tseytlin:1990va,Hull:2006qs,Hull:2006va,Copland:2011wx,Nibbelink:2012jb}.


Another motivation for our work comes from the desire to find and efficiently analyze solvable string backgrounds from the world sheet perspective, \textit{i.e.} from the point of view of the (super)string sigma model. The world sheet theory has long been known of being able to describe non-geometric string backgrounds as well. Examples include T-folds \cite{Hull:2006qs,Hull:2006va},  exotic branes \cite{deBoer:2010ud,Kimura:2013fda,Sakatani:2014hba}, and non-relativistic string theories \cite{Gomis:2000bd,Gomis:2004pw,Gomis:2005pg,Gomis:2005bj,Kim:2007hb,Kim:2007pc,Kim:2008ie,Balasubramanian:2008dm}. The latter ones are of particular interest for applications of the AdS/CFT correspondence to strongly coupled condensed matter systems, which often enjoy non-relativistic (e.g., Galilean, Schr\"odinger, Lifshitz) rather than relativistic symmetries. Historically, the first example of such a sigma model was the non-relativistic closed string theory developed by Gomis and Ooguri in \cite{Gomis:2000bd}, which will be at the heart of this work. We will in particular clarify the non-geometric nature of this string theory background by describing it \textit{via} the DFT sigma model of Ref.~\cite{Lee:2013hma}. We hope that this study will lead to a better general understanding of the properties of non-relativistic vacua in string theory ---similar to how recent studies of non-relativistic gravity \cite{Horava:2009uw,Jensen:2014aia,Hartong:2014oma,Hartong:2014pma,Hartong:2015wxa,Bergshoeff:2014uea,Bergshoeff:2015uaa,Hartong:2015zia} were partially inspired by condensed matter applications---  and further  to additional insight into the structure of gravity itself.

DFT~\cite{Siegel:1993xq,Siegel:1993th,Hull:2009mi,Hull:2009zb,Hohm:2010jy,Hohm:2010pp,Jeon:2012hp} is a formulation of (super)gravity in $D$ dimensions which makes the T-duality symmetries \cite{Buscher:1985kb,Buscher:1987sk,Buscher:1987qj,Giveon:1988tt,Meissner:1991zj} of the (super)string manifest. 
It begins by replacing standard $D$-dimensional spacetime with a $2D$-dimensional \emph{doubled geometry} \cite{Duff:1989tf,Tseytlin:1990nb,Tseytlin:1990va}, equipped with \emph{doubled coordinates} $x^M = (\tilde x_\mu,x^\mu)$.
The tangent space of doubled geometry is equipped with an $\ODD$ invariant metric $\cJ_{MN}$, which in the coordinates $x^M$ takes the off-diagonal form,
\be
\cJ_{MN}=\left(\ba{cc}0 & \delta^\mu_{\phantom\mu\nu} \\
	\delta_\mu^{\phantom\mu\nu} & 0 \ea\right)\,.
\label{cJform}
\ee
%
Because the standard spacetime diffeomorphisms cannot preserve such a structure, the structure group of the manifold must be modified, and hence the action of the Lie derivative as well. The \emph{generalized Lie derivative} preserving $\cJ_{AB}$ takes the form
\be
\hcL_{X}T_{A_{1}\cdots A_{n}}:=X^{B}\partial_{B}T_{A_{1}\cdots A_{n}}+\omega\partial_{B}X^{B}T_{A_{1}\cdots A_{n}}+\sum_{i=1}^{n}(\partial_{A_{i}}X_{B}-\partial_{B}X_{A_{i}})T_{A_{1}\cdots A_{i-1}}{}^{B}{}_{A_{i+1}\cdots  A_{n}}\,.
\label{tcL}
\ee
where $\omega$ denotes the weight of $T_{A_1\cdots A_n}$, and all the $\ODD$ vector indices can be freely raised and lowered by the $\ODD$ invariant constant metric $\cJ_{MN}$. 
The generalized Lie derivative  unifies the Riemannian diffeomorphism and the $B$-field gauge symmetry, similar to the construction used in \textit{Generalized Geometry}~\cite{Gualtieri:2003dx,Hitchin:2004ut,Hitchin:2010qz,
Grana:2008yw,Coimbra:2011nw,Coimbra:2012yy}.
The symmetries of DFT become the generalized diffeomorphisms generated by the generalized Lie derivative, together with the global $\ODD$ rotations comprising the T-duality group~\cite{Siegel:1993th,Gualtieri:2003dx,Grana:2008yw}.


While the doubling of the spacetime coordinates makes T-duality symmetry manifest, it obviously introduces too many spacetime dimensions, and a mechanism is needed to maintain only the physical number of dimensions.
In DFT, this reduction occurs by means of a \textit{section condition}.
One first imposes a second order differential constraint on all objects constructed from local fields, 
\be
\partial_{A}\partial^{A} = 0\,.
\label{seccon0}
\ee
In particular, imposing this condition on the product $\phi_1\phi_2$ of any two DFT fields implies that $\p_A\phi_1\p^A\phi_2 = 0$.
The most general solution to this constraint corresponds to choosing a polarization, or \emph{section condition}, on the tangent space: \emph{i.e.,} we choose a $D$-dimensional subspace that is totally null with respect to $\cJ_{AB}$, and require all fields to have vanishing derivative orthogonal to this subspace.
Such a polarization provides a natural decomposition of the coordinates into two sets as above, of the form $x^{A}=(\tx_{\mu},x^{\nu})$, with ``ordinary'' coordinates, $x^{\nu}$, and T-dual ``winding'' coordinates, $\tx_{\mu}$. 
We will generally fix a section by enforcing that all the fields are independent of the dual winding directions,
\be
{\frac{\partial\Phi}{\,\,\partial \tx_{\mu}}}\equiv0 \,.
\label{sectionSOL}
\ee
This solves the section condition~(\ref{seccon0}), and DFT then reduces to the familiar low energy effective actions of closed (super)string theory. 
During this procedure, supergravity loses manifest $\ODD$ symmetry. 
Note that the group of $\ODD$ transformations changes the section condition while preserving the invariant metric $\cJ_{AB}$.

Technically, the section condition ensures the closure of the algebra of generalized Lie derivatives~(\ref{tcL}), and also that arbitrary functions and their arbitrary derivatives, denoted collectively by $\Phi$, are invariant under the coordinate gauge symmetry \textit{shift} as
\be
\ba{ll}
\Phi(x+\Delta)=\Phi(x)\,,\quad&\quad\Delta^{A}=\phi\partial^{A}\varphi\,.
\ea
\label{aTensorCGS}
\ee
Geometrically it means that~\cite{Park:2013mpa,Lee:2013hma} DFT employs a \textit{doubled-yet-gauged coordinate system} for the description of a $D$-dimensional spacetime: the doubled coordinate system is equipped with an {equivalence relation},
\be
x^{A}~\simeq ~x^{A}+\phi(x) \partial^{A}\varphi (x) 
\label{aCGS}
\ee 
where $\phi,\varphi$ are DFT fields, and each equivalence class, or gauge orbit, represents a single physical point. The diffeomorphism symmetry means an invariance under arbitrary reparametrizations of the gauge orbits.

In DFT, the  geometrical field variables are the $\ODD$ singlet (modified) dilaton, $d$, and the $\ODD$ covariant generalized metric, $\cH_{AB}$.  
The DFT dilaton $d$ is related to the standard string dilaton field $\phi$ by $e^{-2d}=\sqrt{-g}e^{-2\phi}$, making $e^{-2d}$ a scalar density of unit weight, while the generalized metric is defined as a symmetric $\ODD$ element, \emph{i.e.,} it satisfies
\be
\ba{ll}
\cH_{AB}=\cH_{BA}\,,\quad&\quad         \cH_{A}{}^{C}\cH_{B}{}^{D}\cJ_{CD}=\cJ_{AB}\,.
\ea
\ee
By adopting the block off-diagonal form~(\ref{cJform}) of $\cJ_{AB}$, the generalized metric $\cH_{AB}$ also naturally  decomposes into four $D\times D$ blocks.  Now, with respect to the \textit{``canonical" choice of the section}~(\ref{sectionSOL}), if we assume the upper left symmetric block to be non-degenerate and identified with the inverse of a Riemannian metric, it is straightforward to check that the remaining components are determined by a two-form field $B$,
\be
\ba{llll}
\cH_{AB}=\left(\ba{cc}
G^{-1}&-G^{-1}B\\
BG^{-1}&~G-BG^{-1}B
\ea
\right)\,,\quad&~~ G_{\mu\nu}=G_{\nu\mu}\,,\quad&~~ \det(G_{\mu\nu})\neq0\,,
\quad&~~
B_{\mu\nu}=-B_{\nu\mu}\,.
\ea
\label{nondegH}
\ee
Of course, this parametrization is by no means unique. Though it can be a  preferred choice for the canonical section~(\ref{sectionSOL}), one may also perform field redefinitions which change the above identification, \textit{c.f. e.g.~}\cite{Andriot:2013xca}. 

DFT is also well-defined in the cases where the generalized metric $\cH_{AB}$ is an element of $\ODD$, but the upper left block in it is degenerate. In this case, the generalized metric does not allow any Riemannian interpretation. Nevertheless, DFT makes perfect sense even for such a \textit{non-Riemannian} background. The analysis of such backgrounds, in particular the one describing the non-relativistic closed string theory of \cite{Gomis:2000bd}, is the main focus of this work. What is known so far about the behavior of DFT in non-geometric backgrounds (\textit{c.f.}~\cite{Aldazabal:2013sca,Berman:2013eva,Hohm:2013bwa} and references therein) indicates that DFT may provide a novel theoretical framework to formulate string theory,  alternative to, or generalizing,  the conventional Riemannian setup.  In particular the geometric implementation of the section condition as a \textit{doubled-yet-gauged} spacetime \cite{Park:2013mpa} allowed to  formulate a string worldsheet theory \cite{Lee:2013hma} with the doubled spacetime coordinates being dynamical fields  \cite{Hull:2006qs,Hull:2006va,Berman:2013eva}  in which the coordinate gauge symmetry~(\ref{aCGS}) is realized as a usual gauge symmetry on the worldsheet.  This  string  action couples to an arbitrarily curved  generalized metric and dilaton, and is still completely covariant with respect to the coordinate gauge symmetry, DFT diffeomorphisms~(\ref{tcL}),  $\ODD$ T-duality, and the usual  world-sheet diffeomorphisms as well as world-sheet Weyl symmetry. While  it reduces to the standard  Polyakov string action in the above-described Riemannian case~(\ref{nondegH}),  it can also go beyond and in principle describe non-Riemannian backgrounds.  We will make this explicit in this work by finding the non-geometric background which reduces the DFT sigma model of \cite{Lee:2013hma} to the non-relativistic closed string  of \cite{Gomis:2000bd}, and show that the target space DFT equations of motion correctly reproduce the perturbative spectrum of \cite{Gomis:2000bd}, including the winding mode sector. 

The remainder of this work is organized as follows: In section \ref{sec:2}, we start by reviewing the `semi-covariant'  formulation of DFT for the NS-NS sector. We then derive a compact form of the Lagrangian expanded to second order in fluctuations around a generic background, in  terms of completely covariantized differential operators. This is one main result of the paper.  In section \ref{sec:SigmaModels} we introduce the DFT sigma model of \cite{Lee:2013hma}, elaborate on the distinction between geometric and non-geometric (non-Riemannian) backgrounds, and then derive the non-Riemannian    background corresponding to the non-relativistic closed string of \cite{Gomis:2000bd}. In section \ref{sec:4} we  present the other main result of this work,   the spectrum of linear perturbations around the non-Riemannian  DFT background for non-relativistic closed string theory. On the way, we obtain the explicit realization of the Bargmann algebra on the target space DFT manifold, and also present a novel DFT background with Schr\"odinger conformal symmetry. We end with conclusions as well as an outlook on future research directions in section \ref{sec:5}. 
Some technical details of the derivations can be found in  the Appendix.



\section{Linearized Perturbations in Double Field Theory}\label{sec:2}
The analysis of linear perturbations around gravitational backgrounds yields important information about the spectrum of physical excitations, as well as possible pathologies such as tachyonic or ghost instabilities. 
The starting point of our work is the semi-covariant formulation of bosonic (NS-NS sector) DFT~\cite{Jeon:2010rw,Jeon:2011cn}, which we review in section~\ref{sec:DFT}.
We then analyze the fluctuations around a generic background, deriving a compact form of the Lagrangian to second order in fluctuations and expressing the resulting fluctuation equations of motion in terms of \textit{completely covariantized} semi-covariant derivatives. 
Section~\ref{sec:covariance} is devoted to proving the covariance of these equations, and in the course we also derive a novel completely covariant $2^\mathrm{nd}$ order differential operator.

\subsection{Semi-Covariant Formulation of Double Field Theory\label{sec:DFT}}
The semi-covariant formulation of DFT~\cite{Jeon:2010rw,Jeon:2011cn} (c.f. also \cite{Hohm:2011si}) expresses all quantities in terms of the symmetric tensors
\begin{equation}
\ba{ll}
P_{AB} = \frac{1}{2}(\cJ_{AB} + \cH_{AB})\,,
\quad&\quad
\brP_{AB} = \frac{1}{2}(\cJ_{AB} - \cH_{AB})\, .
\ea
\end{equation}
Because the generalized metric $\cH_{AB}$ is an element of $\ODD$, these give rise to complementary projection operators $P^A_{\phantom AB} = \cJ^{AC}P_{CB}$, $\bar P^A{}_{B}=\cJ^{AC}\bar P_{CB}$.
That is, these matrices satisfy
\be
P^{2} = P\,,\qquad
\brP^{2} = \brP\,,\qquad
P\brP=0\,,\qquad 
P+\brP = 
\mathds{1} 
\,.
\label{projection}
\ee
The most geometric way to formulate DFT is to introduce a connection that preserves the relevant geometric structures --- in this case, $P$, $\brP$, and the DFT dilaton $d$.
One can show that these objects are all covariantly constant with respect to the \emph{semi-covariant derivative}~\cite{Jeon:2010rw,Jeon:2011cn}
\be
\na_{C}T_{A_{1}\cdots A_{n}}
=\partial_{C}T_{A_{1}\cdots A_{n}}-\omega\Gamma^{B}{}_{BC} T_{A_{1}\cdots A_{n}}+
\sum_{i=1}^{n}\,\Gamma_{CA_{i}}{}^{B} T_{A_{1}\cdots A_{i-1}BA_{i+1}\cdots A_{n}}\,,
\label{semicovD}
\ee
where the torsionless DFT connection $\Gamma_{CA}^{\phantom{CA}B}$ is defined by
\be
\ba{ll}
\Gamma_{CAB}=&2\left(P\partial_{C}P\brP\right)_{[AB]}
+2\left({{\brP}_{[A}{}^{D}{\brP}_{B]}{}^{E}}-{P_{[A}{}^{D}P_{B]}{}^{E}}\right)\partial_{D}P_{EC}\\
{}&-\textstyle{\frac{4}{D-1}}\left(\brP_{C[A}\brP_{B]}{}^{D}+P_{C[A}P_{B]}{}^{D}\right)\!\left(\partial_{D}d+(P\partial^{E}P\brP)_{[ED]}\right)\,.
\ea
\label{Gamma}
\ee 
``Semi-covariant'' refers to the fact that the semi-covariant derivative of a tensor $T_{A_1\cdots A_n}$ fails to be itself a tensor: its transformation $\delta_X \na_C T_{A_1\cdots A_n}$ under an infinitesimal generalized diffeomorphism $X^A$ is not equal to the generalized Lie derivative $\hat\cL_X \na_C T_{A_1\cdots A_n}$ of \req{tcL}.
Rather, 
\be
(\delta_{X}{-\hcL_{X}})\na_{C}T_{A_{1}\cdots A_{n}}=
\sum_{i=1}^{n}2(\cP{+\brcP})_{CA_{i}}{}^{BFDE}
\partial_{F}\partial_{[D}X_{E]}T_{A_{1}\cdots A_{i-1} BA_{i+1}\cdots A_{n}}\,.
\ee
Here we have introduced six-index projection operators
\be
\ba{ll}
\cP_{CAB}{}^{DEF}:=P_{C}{}^{D}P_{[A}{}^{[E}P_{B]}{}^{F]}+\textstyle{\frac{2}{D-1}}P_{C[A}P_{B]}{}^{[E}P^{F]D}\,,~~&~~{\cP_{CAB}{}^{DEF}\cP_{DEF}{}^{GHI}=\cP_{CAB}{}^{GHI}}\,,\\
\brcP_{CAB}{}^{DEF}:=\brP_{C}{}^{D}\brP_{[A}{}^{[E}\brP_{B]}{}^{F]}+\textstyle{\frac{2}{D-1}}\brP_{C[A}\brP_{B]}{}^{[E}\brP^{F]D}\,,~~&~~{\brcP_{CAB}{}^{DEF}\brcP_{DEF}{}^{GHI}=\brcP_{CAB}{}^{GHI}}\,,
\ea
\label{P6}
\ee
which are traceless ($\cP^{A}{}_{ABDEF}=0$ and $\brcP^{A}{}_{ABDEF}=0$)
and satisfy the symmetry properties
\be
\ba{ll}
\cP_{CABDEF}=\cP_{DEFCAB}=\cP_{C[AB]D[EF]}\,,\quad&\quad
\brcP_{CABDEF}=\brcP_{DEFCAB}=\brcP_{C[AB]D[EF]}\,.
\ea
\label{symP6}
\ee
It is useful to note that $\Gamma_{CAB}$ is the \emph{unique} connection parallelizing $d$, $P$ and $\brP$,
\be
\na_{A}d = -\half e^{2d}\na_{A}(e^{-2d})=\partial_{A}d+\half\Gamma^{B}{}_{BA}=0\,,
\qquad\quad
\na_{A}P_{BC}=0\,,
\qquad\quad
\na_{A}\brP_{BC}=0\,,
\ee
and which also satisfies the vanishing properties $\Gamma_{C(AB)} = \Gamma_{[ABC]}=(\cP+\brcP)_{CAB}{}^{DEF}\Gamma_{DEF}=0$.

As we so-far have not defined a purely covariant derivative (which we will do in section~\ref{sec:covariance}), we must be careful that the objects built from the semi-covariant derivative $\nabla_A$ that appear in actions and equations of motion  satisfy `$\delta_X = \hat\cL_X$' --- \textit{i.e.,} that their transformation under infinitesimal generalized diffeomorphisms is equal to the generalized Lie derivative.
We call such an object \emph{completely covariant}.

The natural Riemann-like curvature constructed from the connection $\Gamma_{CAB}$ is the \emph{semi-covariant curvature},
\be\label{defS}
S_{ABCD}:=\half(R_{ABCD}+R_{CDAB}-\Gamma^{E}{}_{AB}\Gamma_{ECD})\,,
\ee
where $R_{CDAB}$ is the familiar Riemannian expression for the conventional curvature,
\be
R_{CDAB}:=\partial_{A}\Gamma_{BCD}-\partial_{B}\Gamma_{ACD}
+\Gamma_{AC}{}^{E}\Gamma_{BED}-\Gamma_{BC}{}^{E}\Gamma_{AED} \, .
\ee
Using these objects we can give a simple expression for the Lagrangian in NS-NS sector of DFT \cite{Hohm:2010pp,Jeon:2011cn,Jeon:2011sq}:
\be
\cL=\textstyle{\frac{1}{8}}e^{-2d}\left[
(P^{AC}P^{BD}-\brP^{AC}\brP^{BD})S_{ABCD}-2\Lambda\right]\,,
\label{DFTL}
\ee
where $\Lambda$ is the DFT version of the cosmological constant~\cite{Jeon:2011cn}. 
The complete covariance of this Lagrangian will be reviewed  in section \ref{sec:covariance} where we also identify novel completely covariant differential operators.


\subsection{Covariant Analysis of Linearized Perturbations%
\label{sec:perturbations}} 
In this subsection we derive a fully covariant formula for all quadratic fluctuations around an on-shell background of the   DFT Lagrangian~\eqref{DFTL}. 
The resulting formula, \eqref{effectL}, is in particular completely covariant under DFT diffeomorphisms and $\ODD$ transformations. 
It will turn out to be convenient to work with perturbations of the projection operators \eqref{projection}, rather than those of the generalized metric itself. 
These equations will serve as the starting point for our analysis in the following two sections 
 of a non-Riemannian background corresponding to the non-relativistic closed string theory of \cite{Gomis:2000bd}.
We begin with the DFT equations of motion, which are obtained from the (formal) variational principle for the Lagrangian \req{DFTL} by varying $P\rightarrow P + \delta P$ and $\brP \rightarrow \brP + \delta\brP$.
Because $P$ and $\brP$ are constrained by the quadratic relations~(\ref{projection}), the first order variations are constrained to satisfy 
\be
\delta P + \delta\brP = 0 \,,
\qquad
P\delta P=\delta P\brP\,,
\qquad
\delta P P = \brP \delta P \,,
\ee
such that
\be
\delta P=P\delta P\brP+\brP\delta P P\,.
\label{deltaPconstraint}
\ee
The variation of the semi-covariant curvature \eqref{defS} takes the form
\be
\delta S_{ABCD}
=\na_{[A}\delta\Gamma_{B]CD}+\na_{[C}\delta\Gamma_{D]AB}\,,
\label{deltaS}
\ee
where the variation of the DFT connection \eqref{Gamma} is given explicitly by
\be
\ba{ll}
{\delta\Gamma}_{CAB}=&2P_{[A}{}^{D}\brP_{B]}{}^{E}\DO_{C}\delta P_{DE}+2(\brP_{[A}{}^{D}\brP_{B]}{}^{E}-P_{[A}{}^{D}P_{B]}{}^{E})\DO_{D}\delta P_{EC}\\
{}&-\textstyle{\frac{4}{D-1}}(\brP_{C[A}\brP_{B]}{}^{D}+P_{C[A}P_{B]}{}^{D})(\partial_{D}\delta d+P_{E[G}\DO^{G}\delta P^{E}{}_{D]})\\
{}&-\Gamma_{FDE\,}\delta(\cP+\brcP)_{CAB}{}^{FDE}\,.
\ea
\label{deltaGamma}
\ee
The variation of the Lagrangian~\req{DFTL} can now be written
\be
\delta\cL = -2\cL\,\delta d+\quarter e^{-2d}S_{AB}\,\delta P^{AB}+\quarter\partial_{A}
\left[e^{-2d}(P^{AC}P^{BD}-\brP^{AC}\brP^{BD})\delta\Gamma_{BCD}\right]\,,
\label{deltaL}
\ee
where we have defined $S_{AB}=S_{ACB}{}^{C}$.
We thus obtain the two DFT equations of motion%
\footnote{Note the following equivalence due to the projection property, 
\[
\ba{lll}
(PS\brP)_{(AB)} = 0 \quad&\Longleftrightarrow&\quad (PS\brP)_{AB} = 0 \,.
\ea
\]}
\be
\ba{ll}
(PS\brP)_{AB}=P_{A}{}^{C}\brP_{B}{}^{D}S_{CD} \equiv 0\,, \quad&\quad
\cL=\textstyle{\frac{1}{8}}e^{-2d}\left[
(P^{AC}P^{BD}-\brP^{AC}\brP^{BD})S_{ABCD}-2\Lambda\right]\equiv 0 \,.
\ea
\label{DFT-EOM}
\ee
Hereafter, the equivalence symbol, `\,$\equiv$\,' means the on-shell relation.
The equations of motion for the  quadratic fluctuations around an on-shell background are found by taking the on-shell variations of the equations of motion~\req{DFT-EOM}, which are
\begin{align}
\delta\cL
&\equiv\quarter
e^{-2d}(P^{AC}P^{BD}-\brP^{AC}\brP^{BD})\na_{A}\delta\Gamma_{BCD}
\nonumber\\
{}& = e^{-2d}\left[\half(P^{AB}-\brP^{AB})\na_{A}\partial_{B}\delta d-\quarter\na_{A}\na_{B}\delta P^{AB}\right]\,,
\label{deltaL2}
\end{align}
and
\begin{align}
\delta(PS\brP)_{AB}  & 
= 2P_{A}{}^{C}\brP_{B}{}^{D}\na_{C}\partial_{D}\delta d
+\half(P_{A}{}^{C}\brDelta_{B}{}^{D}-\Delta_{A}{}^{C}\brP_{B}{}^{D})\delta P_{CD}-2(PS\brP\delta PP)_{(AB)} 
\nonumber\\
{}& 
\equiv 2P_{A}{}^{C}\brP_{B}{}^{D}\na_{C}\partial_{D}\delta d
+\half(P_{A}{}^{C}\brDelta_{B}{}^{D}-\Delta_{A}{}^{C}\brP_{B}{}^{D})\delta P_{CD}\,,
\label{deltaRicci}
\end{align}
where we have defined a pair of second order (semi-covariant) differential operators, 
\be
\ba{l}
\Delta_{A}{}^{B}:=P_{A}{}^{B}P^{CD}\na_{C}\na_{D}  -2P_{A}{}^{D}P^{BC}(\na_{C}\na_{D}-S_{CD})\,,\\
\brDelta_{A}{}^{B}:=\brP_{A}{}^{B}\brP^{CD}\na_{C}\na_{D}-2\brP_{A}{}^{D}\brP^{BC}(\na_{C}\na_{D}-S_{CD})\,.
\ea
\label{DeltabrDelta}
\ee
In summary, we have the following results:
\begin{framed}
\noindent
The DFT fluctuations $\delta d$ and $\delta P_{AB}$ satisfy the equations of motion:
\be
(P^{AB}-\brP^{AB})\na_{A}\partial_{B}\delta d-\half\na_{A}\na_{B}\delta P^{AB}\equiv 0\,,
\label{EOMf1}
\ee
\be
P_{A}{}^{C}\brP_{B}{}^{D}\na_{C}\partial_{D}\delta d+\quarter
(P_{A}{}^{C}\brDelta_{B}{}^{D}-\Delta_{A}{}^{C}\brP_{B}{}^{D})\delta P_{CD}\equiv 0\,.
\label{EOMf2}
\ee
These two relations can be also  derived from the following effective Lagrangian for the fluctuations around a given on-shell background,
\be
\cL_{{\rm eff.}}:=e^{-2d}\left[\half(P-\brP)^{AB}\partial_{A}\delta d\,\partial_{B}\delta d-\half\partial_{A}\delta d\,\na_{B}\delta P^{AB}+\textstyle{\frac{1}{8}}\delta P^{AB}(\brDelta_{A}{}^{C}P_{B}{}^{D}-\Delta_{A}{}^{C}\brP_{B}{}^{D})\delta P_{CD}\right].\quad~
\label{effectL}
\ee
\textit{Gauge symmetries.}  As we show in the next section, the expressions (\ref{EOMf1}-\ref{effectL}), as off-shell quantities, transform completely covariantly under DFT diffeomorphisms, given in terms of the generalized Lie derivative acting on each field,
\be
\ba{llll}
\delta_{X}d=\hcL_{X}d\,,\quad&\delta_X P_{AB}=\hcL_{X}P_{AB}\,,\quad&
\delta_{X}(\delta d)=\hcL_{X}\delta d=X^{A}\partial_{A}\delta d\,,\quad&\delta_{X}(\delta P_{AB})=\hcL_{X}\delta P_{AB}\,.
\ea
\ee
Furthermore, the two equations of motion  (\ref{EOMf1}), (\ref{EOMf2}) are also covariant under the linearized diffeomorphism transformation $\delta_Y$, which acts according to $\delta_{Y} d=\delta_{Y} P_{AB}=0$ and
\be
\delta_{Y}(\delta d)=\hcL_{Y}d=-\half e^{2d}\hcL_{Y}(e^{-2d})=Y^{A}\partial_{A}d-\half \partial_{A}Y^{A}\,,\quad
\quad\delta_{Y}(\delta P_{AB})=\hcL_{Y}P_{AB}\,.
\ee
\end{framed}
\noindent
Note that, although the cosmological constant, $\Lambda$, does not appear explicitly above, its effect is included in the dependence on the on-shell background, which solves (\ref{DFT-EOM}). 
The effective Lagrangian~\req{effectL} can also be derived directly by expanding the original DFT Lagrangian~(\ref{DFTL}) to second order in the fluctuations. To do so,  one must take care with the constraints~\req{projection}, which imply that the quadratic variation of $P,\brP$ are non-zero. 
Instead, the variations take the form
\be
d\;\;\mapsto\;\; d+\delta d\,,\quad
\qquad 
P\;\;\mapsto\;\; P+\delta P+\brP(\delta P)^{2}\brP-P(\delta P)^{2} P \,.
\ee
As shown explicitly in Appendix \ref{app:A}, the effective Lagrangian above~(\ref{effectL}) then agrees (up to total derivatives) on-shell  with the second order variation of the original DFT Lagrangian ,
\be
\cL_{{\rm eff.}} = \half\delta^{2}\cL\,.
\ee
%

\subsection{Covariance of the Linearized Equations, and  new 2$^\mathrm{nd}$ Order Differential Operators%
\label{sec:covariance}}  

In this subsection,  we check  the full DFT diffeomorphism   covariance of the  fluctuation equations,  \req{EOMf1}, \req{EOMf2}, explicitly. 
We start by recalling from \cite{Jeon:2011cn} that the six-index projectors~\req{P6} govern the  anomalous (\textit{i.e.}  non-covariant) terms of  the  semi-covariant derivative and curvature  under generalized diffeomorphisms:
\begin{align}
& (\delta_{X}{-\hcL_{X}})\Gamma_{CAB}= 2\big[(\cP{+\brcP})_{CAB}{}^{FDE}-\delta_{C}^{~F}\delta_{A}^{~D}\delta_{B}^{~E}\big]
\partial_{F}\partial_{[D}X_{E]}\,, 
\nonumber\\
& \dis{(\delta_{X}{-\hcL_{X}})\na_{C}T_{A_{1}\cdots A_{n}}=
\sum_{i=1}^{n}2(\cP{+\brcP})_{CA_{i}}{}^{BFDE}
\partial_{F}\partial_{[D}X_{E]}T_{A_{1}\cdots A_{i-1} BA_{i+1}\cdots A_{n}}\,,} 
\label{anomalous}\\
& (\delta_{X}-\hcL_{X})S_{ABCD}= 
2\na_{[A}\left((\cP{+\brcP})_{B][CD]}{}^{EFG}\partial_{E}\partial_{[F}X_{G]}\right)
+2\na_{[C}\left((\cP{+\brcP})_{D][AB]}{}^{EFG}\partial_{E}\partial_{[F}X_{G]}\right)\,.
\nonumber
\end{align}
From the definition \req{P6} it can be seen easily that any contraction of $P$ with $\brcP$, or of $\brP$ with $\cP$ vanishes.
It follows then immediately  that, by applying appropriate projection operators, one can construct completely covariant tensors from the semi-covariant derivative, for example \cite{Jeon:2011cn}:
\be
\ba{ll}
P_{C}{}^{D}{\brP}_{A_{1}}{}^{B_{1}}\cdots{\brP}_{A_{n}}{}^{B_{n}}
\DO_{D}T_{B_{1}\cdots B_{n}}\,,~&~
{\brP}_{C}{}^{D}P_{A_{1}}{}^{B_{1}}\cdots P_{A_{n}}{}^{B_{n}}
\DO_{D}T_{B_{1}\cdots B_{n}}\,,\\
P^{AB}{\brP}_{C_{1}}{}^{D_{1}}\cdots{\brP}_{C_{n}}{}^{D_{n}}\DO_{A}T_{BD_{1}\cdots D_{n}}\,,~&~
\brP^{AB}{P}_{C_{1}}{}^{D_{1}}\cdots{P}_{C_{n}}{}^{D_{n}}\DO_{A}T_{BD_{1}\cdots D_{n}}\,,\\
P^{AB}{\brP}_{C_{1}}{}^{D_{1}}\cdots{\brP}_{C_{n}}{}^{D_{n}}
\DO_{A}\DO_{B}T_{D_{1}\cdots D_{n}}\,,~&~
{\brP}^{AB}P_{C_{1}}{}^{D_{1}}\cdots P_{C_{n}}{}^{D_{n}}
\DO_{A}\DO_{B}T_{D_{1}\cdots D_{n}}\,.
\ea
\label{covT}
\ee
Similarly, the following curvatures are completely covariant:
\be
(PS\brP)_{AB}=P_{A}{}^{C}\brP_{B}{}^{D}S_{CD}\,,
\qquad\quad
(P^{AC}P^{BD}-\brP^{AC}\brP^{BD})S_{ABCD}\,.
\label{aRicciScalar}
\ee
It is worthwhile to note also the identities  for the completely covariant scalar curvature,\footnote{
These identities may be used to simplify \eqref{DFTL} by showing that the first two terms are equal.  However, in the full order supersymmetric extensions of  DFT~\cite{Jeon:2011sq,Jeon:2012hp} the connection $\Gamma_{CAB}$ of equation~(\ref{Gamma}) is no longer torsion-free, and the identity~(\ref{idS}) no longer holds. In this case, \req{aRicciScalar} is the correct covariant object.}
\be
P^{AC}P^{BD}S_{ABCD}\seceq P^{AB}S_{AB}\seceq-\brP^{AC}\brP^{BD}S_{ABCD}\seceq-\brP^{AB}S_{AB}\,.
\label{idS}
\ee

Next we define a pair of second order (semi-covariant)  differential operators,
\be\label{complcov}
\ba{l}
\fD_{A}{}^{B}:=(P_{A}{}^{B}P^{CD}-2P_{A}{}^{D}P^{BC})(\na_{C}\na_{D}-S_{CD})\,,\\
\brfD_{A}{}^{B}:=(\brP_{xA}{}^{B}\brP^{CD}-2\brP_{A}{}^{D}\brP^{BC})(\na_{C}\na_{D}-S_{CD})\,.
\ea
\ee
These operators  are closely related to  $\Delta_{A}{}^{B}$, $\brDelta_{A}{}^{B}$~(\ref{DeltabrDelta}) simply by the completely covariant scalar curvature,
\be
\ba{ll}
\fD_{A}{}^{B}=\Delta_{A}{}^{B}-P_{A}{}^{B}P^{CD}S_{CD}\,,\quad&\quad
\brfD_{A}{}^{B}=\brDelta_{A}{}^{B}-\brP_{A}{}^{B}\brP^{CD}S_{CD}\,.
\ea
\label{diffoprel}
\ee
The successive application of  (\ref{anomalous}) yields for the semi-covariant second derivative,
\begin{align}
(\delta_{X}{-\hcL_{X}})\na_{B}\na_{C}T_{A_{1}\cdots A_{n}} = 
& \;
2(\cP{+\brcP})_{BC}{}^{GFDE}
\partial_{F}\partial_{[D}X_{E]}\na_{G}T_{A_{1}\cdots A_{n}} 
\label{noncov}\\
{}&+
\sum_{i=1}^{n}2(\cP{+\brcP})_{BA_{i}}{}^{GFDE}
\partial_{F}\partial_{[D}X_{E]}\na_{C}T_{A_{1}\cdots A_{i-1} GA_{i+1}\cdots A_{n}}
\nonumber\\
{}&+
\sum_{i=1}^{n}2(\cP{+\brcP})_{CA_{i}}{}^{GFDE}
\partial_{F}\partial_{[D}X_{E]}\na_{B}T_{A_{1}\cdots A_{i-1} GA_{i+1}\cdots A_{n}}
\nonumber\\
{}&+
\sum_{i=1}^{n}2(\cP{+\brcP})_{CA_{i}}{}^{GFDE}
\left(\na_{B}\partial_{F}\partial_{[D}X_{E]}\right)T_{A_{1}\cdots A_{i-1} GA_{i+1}\cdots A_{n}}\,,
\nonumber
\end{align}
which in turn can be used to show that the following contractions of the new operators~\eqref{diffoprel}  with projectors are  \textit{completely covariant second order derivatives,}
\be
\ba{ll}
\fD_{A}{}^{C}\brP_{B_{1}}{}^{D_{1}}\cdots\brP_{B_{n}}{}^{D_{n}}T_{CD_{1}\cdots D_{n}}\,,\quad&\quad
\brfD_{A}{}^{C}P_{B_{1}}{}^{D_{1}}\cdots P_{B_{n}}{}^{D_{n}}T_{CD_{1}\cdots D_{n}}\,,
\ea
\label{comp1}
\ee
\be
\ba{ll}
\Delta_{A}{}^{C}\brP_{B_{1}}{}^{D_{1}}\cdots\brP_{B_{n}}{}^{D_{n}}T_{CD_{1}\cdots D_{n}}\,,\quad&\quad
\brDelta_{A}{}^{C}P_{B_{1}}{}^{D_{1}}\cdots P_{B_{n}}{}^{D_{n}}T_{CD_{1}\cdots D_{n}}\,.
\ea
\label{comp2}
\ee
To the best of our knowledge these operators have not appeared in the literature before. 
They complement the known list of completely covariant tensorial differential operators (\ref{covT}) \cite{Jeon:2011cn} (see also  the appendix of \cite{Jeon:2012kd} for ``Dirac'' operators).

From our list of completely covariant objects (\ref{covT}), \eqref{comp1}, \eqref{comp2}, together with the constraints (\ref{deltaPconstraint}) on $\delta P$, each term in (\ref{deltaL2}) and (\ref{deltaRicci}), 
\be
\ba{lll}
P^{AB}\na_{A}\partial_{B}\delta d\,,\quad&\quad
\brP^{AB}\na_{A}\partial_{B}\delta d\,,\quad&\quad
\na_{A}\na_{B}\delta P^{AB}\,,\\
P_{A}{}^{C}\brP_{B}{}^{D}\na_{C}\partial_{D}\delta d\,,\quad&\quad
P_{A}{}^{C}\brDelta_{B}{}^{D}\delta P_{CD}\,,\quad&\quad
\Delta_{A}{}^{C}\brP_{B}{}^{D}\delta P_{CD}\,,
\ea
\ee 
can be seen to be completely covariant under both generalized diffeomorphisms and $\ODD$ rotations.
This establishes the complete covariance of the equations of motion for the fluctuations.

\newpage
\section{Non-Riemannian Sigma Models and Non-Relativistic Closed Strings%
\label{sec:SigmaModels}}
Gomis and Ooguri \cite{Gomis:2000bd} showed that there exist double scaling limits of the closed string whose symmetries are not the Lorentz group, but rather the non-relativistic group of Galilean transformations.
While the Gomis-Ooguri string is described by a sigma model, it cannot be formulated purely in terms of the target space fields, needing additional worldsheet variables to construct the space of physical states.
This makes it unlikely that its low energy excitations can be described simply in terms of general relativity.

In this section we review the setup, limit and sigma model of the Galilean invariant string theory of \cite{Gomis:2000bd}.
We then show that this non-Riemannian limit has a natural embedding into Double Field Theory, making Double Field Theory a natural candidate for describing the (generalized) geometric structure of the Gomis-Ooguri string.

\subsection{Non-Relativistic Limit of Closed String Theory and the Gomis-Ooguri Sigma Model\label{ssec:NRCS}}

The non-relativistic limit of \cite{Gomis:2000bd} starts with a scaling limit of relativistic string theory incorporating components of a $p$-form gauge potential.
States uncharged under this potential decouple, while charged states acquire a non-relativistic dispersion relation. 
While such limits exist for both open and closed string theories, as in \cite{Gomis:2000bd} we will focus on the non-relativistic closed string limit employing the Neveu-Schwarz $B$ field.

Consider a closed string on flat space, $g_{\mu\nu}=\eta_{\mu\nu}$, winding around a compact spatial circle of radius $R$ denoted by $x^1$, with non-vanishing $B_{01}=B$ along this compact direction. 
We can further naturally introduce a speed of light $c$ by splitting the spacetime coordinates $x^\mu=(x^\alpha,x^i)$ ($\alpha=0,1$ and $i=2,\ldots,D-1$) and rescaling the metric $g_{\alpha\beta}$ in the $x^\alpha$ sector equal to ${c^2}\eta_{\alpha\beta}$.
The dispersion relation of a closed string winding $w$ times around the circle is then
\be
\label{eq:421}
\frac{1}{c^2}(E + \frac{wRB}{\alpha'})^2 = k^2 + c^2\left(\frac{wR}{\alpha'}\right)^2 + \frac{1}{c^2}\left(\frac{n}{R}\right)^2 + \frac{2}{\alpha'}(N+\tilde N-2) \, ,
\ee
together with the level-matching condition $wn=N-\tilde N$.
Here $n$ denotes the $x^1$ momentum quantum, $E$ the energy, and $k^i$ the momentum in the $x^i$ ($i=2,\ldots, D-1$) directions.
$N$ and $\tilde N$ are the stringy excitation numbers in the left and right moving sector, respectively. 

While we cannot take the $c\to\infty$ limit as is, if we also take $B=c^2-\mu$, with $\mu$ finite as $c\to\infty$, then we obtain a finite dispersion relation
\begin{equation}
E = \mu \frac{wR}{\alpha'} + \frac{\alpha' k^2}{2wR} + \frac{N+\tilde N-2}{wR} \; .
\label{eq:GO-spectrum}
\end{equation}
This dispersion relation is precisely that for a Galilean particle with mass and charge $wR/\alpha'$ and chemical potential $\mu$, together with an intrinsic contribution from the string oscillators.
Demanding positive energy states selects positive windings, $w>0$. 

The essence of the limit is its near-criticality, with the divergence due to the winding mode mass exactly canceling the divergence due to the winding mode charge. 

To obtain a non-relativistic limit of the worldsheet theory, we once again rescale the metric in the $x^\alpha$ sector by a constant, $g_{\alpha\beta}=G\eta_{\alpha\beta}$, and take $g_{ij}=\delta_{ij}$ in the other directions.
Working in terms of light-cone coordinates $\gamma= X^0 + X^1$, $\bar\gamma = - X^0 + X^1$, the (Euclidean) string sigma model in conformal gauge has the form
\begin{equation}
S = \frac{1}{4\pi\alpha'}\int d^2z\left( (G - B)\p\gamma\bar\p\bar\gamma + (G+B)\p\bar\gamma\bar\p\gamma + 2\p X^i \bar\p X^i \right).
\label{eq:starting-sigma-model} 
\end{equation}
We wish to take the limit $G\to\infty$.
While this is a singular limit of \req{eq:starting-sigma-model}, introducing Lagrange multipliers $\beta,\bar\beta$ we obtain the equivalent action
\begin{equation}
S = \frac{1}{2\pi\alpha'}\int d^2z\left( \beta\bar\p\gamma + \bar\beta\p\bar\gamma - \frac{2}{G+B}\beta\bar\beta + \frac{1}{2}(G-B)\p\gamma\bar\p\bar\gamma + \p X^i\bar\p X^j \right) .
\end{equation}
Setting $B = G - \mu$, the $G\to\infty$ is now straightforward, and we obtain the worldsheet action
\begin{equation}
S = \frac{1}{2\pi\alpha'}\int d^2z\left( \beta\bar\p\gamma + \bar\beta\p\bar\gamma + \frac{\mu}{2}\p\gamma\bar\p\bar\gamma + \p X^i\bar\p X^i \right) .
\label{eq:GO-action}
\end{equation}
The third term in the action contributes only contact terms to correlation functions, but it modifies the energy in the winding sectors, shifting the spectrum by an amount proportional to the winding number $w$.
$\mu$ can therefore be understood as a residual chemical potential for the winding number.

%
%
The string spectrum for this sigma model was calculated in~\cite{Gomis:2000bd} and in our notation takes precisely the form of equation~\req{eq:GO-spectrum}.
In particular, there are no excitations at winding number zero.
Note that to go from our notation to that of \cite{Gomis:2000bd} one should make the following replacements:
\be
\beta \to \alpha'\beta_\mathrm{GO}
\qquad
\bar\beta \to \alpha'\bar\beta_{\mathrm{GO}}
\qquad
\mu \to \frac{1}{2}
\qquad
\alpha' \to \alpha'_\mathrm{eff}
\,.
\ee
In particular, the quantity $\mu$ takes the fixed value $\half$ in \cite{Gomis:2000bd}.

Finally, it was shown in \cite{Seiberg:2000ms,Gopakumar:2000na,Gomis:2000bd} that we must take a simultaneous strong-coupling limit, which in terms of our parameter $G$ takes the form $g_s = \sqrt{G}g_0$, with $g_0$ held constant~\cite{Seiberg:2000ms,Gopakumar:2000na,Gomis:2000bd}.
$g_0$ is the natural expansion parameter defining perturbation theory in string loops.

The Gomis-Ooguri sigma model crucially involves the worldsheet variables $\beta,\bar\beta$, yet these have no straightforward interpretation as geometric objects within standard Riemannian geometry.
We now turn to establishing an interpretation within the geometric framework provided by the sigma model formulation of Double Field Theory.

\subsection{A Double Field Theory Sigma Model\label{secString}}
As Double Field Theory is intended to make manifest T-duality, an inherently stringy symmetry, it should come as no surprise that the worldsheet theory of the string can be modified in such a way that the geometric structures of Double Field Theory become manifest.
Such a description was found in \cite{Lee:2013hma}, which gave a completely covariant  worldsheet description of a string propagating in the doubled-yet-gauged spacetime described in the introduction.
Recall that the doubled coordinate space is equipped with the equivalence relation (gauge symmetry)~(\ref{aCGS}),  
\be
x^{M}~\sim~ x'{}^{M} = x^{M}+\phi\partial^{M}\varphi \,,
\label{equivCGS}
\ee
where $\phi$ and $\varphi$ are arbitrary fields satisfying the section condition.
Clearly, the usual differential one form $\rd x^{M}$ is not invariant under this gauge transformation. Furthermore, it is not a tensor: its transformation under DFT diffeomorphisms is not given by the generalized Lie derivative~(\ref{tcL}). 
To construct a gauge-invariant one form requires a corresponding gauge connection,
\be
Dx^{M}:=\rd x^{M}-\cA^{M}\,.
\ee
As the connection is  a `derivative index-valued vector' which can be written as  (the sum of) the form~(\ref{equivCGS}), `\,$\phi\partial^{M}\varphi$\,' (see also \eqref{sectionf}), it is natural to require the gauge potential to satisfy its own `section condition' (\textit{c.f.~}\cite{Hohm:2013jma}),\footnote{It is also  worth while to note that, if we regard  $\cA^{M}$ itself as a DFT field so that $\p^N\!\cA^M\p_N = 0$ is satisfied then  we may obtain a  suggestive form of a ``gauged section condition'' like  $
(\partial_{M}+\cA_{M})(\partial^{M}+\cA^{M}) = 0$.} 
\be
\ba{ll}
\cA^{M}\partial_{M} = 0\,,\quad&\quad\cA^{M}\cA_{M}=0\,. 
\ea
\label{gSECCON}
\ee
Thanks to the gauge connection, $Dx^{M}$ is a DFT covariant vector.  Under the coordinate gauge symmetry,   the transformation of $\cA^M$ is chosen such that $Dx^{M}$ remains invariant,
\be
\cA^{M}~~
\longrightarrow
~~\cA^{\prime M}=\cA^{M}+\rd(\Phi_1\partial^{M}\Phi_2)\,, 
%
\label{GD}
\ee
Further, its transformation under DFT diffeomorphisms matches the action of the generalized Lie derivative; see \cite{Lee:2013hma} for details, especially the transformation rules of the gauge potential. 

On a string worldsheet $\Sigma$ with coordinates $\sigma^{a}$ ($a=0,1$), the doubled target spacetime  coordinates and the gauge connection become worldsheet fields, $X^{M}(\sigma)$ and $\cA_{a}^{M}(\sigma)$, so that $X:\Sigma \to {\mathbb{R}}^{D+D}$ and
\be
\rdA X^{M}=\rd\sigma^{a}\rdA_{a}X^{M}=\rd\sigma^{a}(\partial_{a}X^{M}-\cA_{a}^{M})\,.
\ee
The coordinate gauge symmetry is then realized literally as one of the local symmetries in the DFT worldsheet action proposed in~\cite{Lee:2013hma}:
\be
\cS={\textstyle{\frac{1}{4\pi\alpha^{\prime}}}}{\dis{\int_{\Sigma}}}\rd^{2}\sigma~\cL\,,
\qquad
\cL=-\half\sqrt{-h}h^{ab}\rdA_{a}X^{M}\rdA_{b}X^{N}\cH_{MN}(X)-\epsilon^{ab}\rdA_{a}X^{M}\cA_{bM}\,,
\label{Lagrangian}
\ee
where the string tension is halved~\cite{Hull:2006va} and the gauge connection, $\cA_{a}^{M}$, is taken as an auxiliary field to be integrated over in the worldsheet path integral.
This action describes a string propagating in a doubled-yet-gauged spacetime given by a generalized metric $\cH=P-\brP$ which satisfies the section condition. In addition to the coordinate gauge symmetry, the action~\req{Lagrangian} is also invariant under $\ODD$ T-duality rotations as well as under DFT diffeomorphisms~\cite{Lee:2013hma}. 

By fixing the section as $\frac{\partial~~\,}{\partial \tx_{\mu}} = 0$ (\ref{sectionSOL}),  the derivative-index-valued gauge potential assumes the concrete form
\be
 \cA^{M}= A_{\lambda}\partial^{M}x^{\lambda}=(A_{\mu},0)\,,
\label{sectionf}
\ee
which obviously solves the `gauged section condition'~(\ref{gSECCON}).
It follows that
\be
D_{a}X^{M}=(\partial_{a}\tX_{\mu}-A_{a\mu}\,,\,\partial_{a}X^{\mu})\,.
\ee
Note that the sigma model retains the local gauge symmetry
\be
\delta \cA_{a\mu} = \p_a\lambda_\mu\,,
\qquad
\delta \tX_\mu = \lambda_\mu \,,
\label{eq:gauge-shift-symmetry}
\ee
where the periodicity of large gauge transformations is fixed by the periodicity (if any) of $\tilde X_\mu$.

With respect to this choice of the section, the generalized metric can then be  classified into two types: 
\begin{itemize}
\item The \underline{Riemannian} case is given by a generalized metric of which the upper left $D\times D$ block is \textit{non-degenerate}, such that it admits the well-known parametrization in terms of the $D$-dimensional `Riemannian metric' and the  Kalb-Ramond $B$-field, as shown in (\ref{nondegH}).   
In this case, after integrating out the auxiliary gauge connection we recover the standard string sigma model.

Further,  with the assumed non-degeneracy of the Riemannian metric,  
the equation of motion of the gauge connection implies the duality relation between $X^\mu$ and $\tX_\mu$ on the doubled-yet-gauged target spacetime:
\be
g^{\mu\nu}\rdA_{a}\tX_\nu - B^{\mu}_{\phantom{\mu}\nu}\p_{a}X^\nu
+\textstyle{\frac{1}{\sqrt{-h}}}\epsilon_{a}{}^{b}\p_{b}X^{\mu}=0\,.
\label{eq:self-duality}
\ee

\item A \underline{Non-Riemannian} generalized metric is characterized by a \textit{degenerate} upper left $D\times D$ block, and so does not admit a Riemannian interpretation with respect to the section choice~(\ref{sectionSOL})~\cite{Lee:2013hma} (also \emph{c.f.}~\cite{Garcia-Fernandez:2013gja}). The equation of motion of the gauge connection does not generically imply the standard duality relation~(\ref{eq:self-duality}), but a modified duality relation, which we consider in more detail in the next section.

\end{itemize}

\subsection{Non-Riemannian Backgrounds of DFT\label{ssec:GalileanSigmaModel}}
To better understand the ``non-Riemannian'' case, let us review a particular class of such Double Field Theory backgrounds obtained in \cite{Lee:2013hma} by T-duality, and their realization by the sigma model \eqref{Lagrangian}.

As in \cite{Lee:2013hma}, we start with a generalization of the exact solution of supergravity obtained in \cite{Dabholkar:1990yf} corresponding to a macroscopic fundamental string geometry in ten dimensions
\be
\ba{rclrcl}
ds^2 &=& \multicolumn{4}{l}{f^{-1}(-dt^2 + (dx^1)^2) + (dx^2)^2 + \dots + (dx^9)^2\,,} \\
B &=& (f^{-1} - \hat{c}) dt \wedge dx^1\,,\qquad\quad & e^{-2\phi} &=& f\, e^{-2\phi_0} \,,\\ 
f &=& 1+\frac{Q}{r^{6}}\,,\qquad & r^{2} &=& \sum_{a=2}^{9} (x^{a})^{2}\,,
\ea
\ee
where $Q$ is the number of string quanta, and $\phi_0$ and $\hat{c}$ are constants. 
Note that $1-\hat{c}$ is the chemical potential for winding charge; as a result,
when $x^1$ is compactified, we must take $\hat{c}\in(0,2)$, otherwise the background 
is unstable to the spontaneous condensation of winding strings.
(If $x^1$ is non-compact then $\hat{c}$ is pure gauge and can be removed.)
This solution splits the spacetime into the directions parallel and transverse to the string, transforming under an $\SO(1,1) \times \SO(8)$ subgroup of $\SO(1,9)$.
In the following, the greek letters $\alpha,\beta,\gamma,\delta, \cdots$ denote the  Minkowskian $\SO(1,1)$ vector indices subject to the flat metric $\eta_{\alpha\beta}=\mbox{diag}(-+)$, and the roman letters $i,j,k,\ell\cdots$ are for the Euclidean $\SO(8)$ vector indices with flat metric $\delta_{ij}$.  The doubled coordinate $X^M$ splits into $(\tilde x_\alpha,\tilde x_i,x^\alpha,x^i)$, and so the generalized metric decomposes into sixteen blocks, as $(2+8+2+8)\times (2+8+2+8)$.
Further, with the $2\times 2$ anti-symmetric Levi-Civita symbol, $\cE_{\alpha\beta}=-\cE_{\beta\alpha}$,  $\cE_{01}=+1$, we set 
\be
\cE^{\alpha}{}_{\beta} = \eta^{\alpha\gamma}\cE_{\gamma \beta}=  -\cE_{\beta}{}^{\alpha}=-\cE_{\beta\delta}\eta^{\delta\alpha}\,,
\ee
which satisfies
\be
\cE^{\alpha}{}_{\beta}\cE^{\beta}{}_{\gamma}=\delta^{\alpha}{}_{\gamma}\,.
\ee
%
Now we perform an $\ODD$ rotation (\emph{i.e.,} a T-duality) which exchanges the $(t,x^{1})$ and $(\tilde{t},\tx_{1})$ planes,%
\footnote{
Note that the $\ODD$ rotation here may not correspond to the traditional T-duality rotation. In backgrounds with isometries, we choose the coordinates, $x^A=(\tilde{x}_\alpha,\tilde{x}_i,x^\alpha,x^i)$, such that the background fields are independent of $\tilde{x}_i$ and $x^I=(\tilde{x}_\alpha,x^\beta)$. In such backgrounds, a global $\ODD$ rotation, $\cH_{AB} \rightarrow O_A{}^C\,O_B{}^D\, \cH_{CD}$ with $O_A{}^B=\bigl(\begin{smallmatrix} 1 & 0 \cr 0 & O_I{}^J \end{smallmatrix}\bigr) \in \ODD$ (keeping the coordinates fixed), transforms the equation of motion of DFT covariantly.  We used this rotation as a solution generating method. For discussions  on T-duality along the temporal direction, see \textit{e.g.~}\cite{Moore:1993zc,Malek:2013sp} and also on related subtle issues, see \cite{Cederwall:2014kxa,Cederwall:2014opa}.\label{footR}}
\be
\ba{ll}
\cH_{AB}\quad\longrightarrow\quad\cO_{A}{}^{C}\cO_{B}{}^{D}\cH_{CD}\,,\quad&\quad
\cO_{A}{}^{B}=\left(\ba{cccc}
0 & 0 & \eta^{\alpha\beta} & 0 \\
0 & \delta^{i}{}_{j}&0&0\\
\eta_{\alpha\beta}&0&0&0\\
0 & 0 & 0 & \delta_{i}{}^{j}
\ea
\right)\,
\ea
\ee
to obtain a new generalized metric of the form
\be
\cH_{MN} =\left(\ba{cccc}
\hat{c}(2-\hat{c}f) \eta^{\alpha\beta}&0&   (1-\hat{c} f) \cE^{\alpha}{}_{\beta}&0\\
0&\delta^{ij}&0&0\\
-(1 - \hat{c} f)\cE_{\alpha}{}^{\beta} &0 & f\eta_{\alpha\beta}&0\\
0&0&0&\delta_{ij}\ea\right)\, .
\label{dualgeom}
\ee
corresponding to the geometric configuration
\be
\ba{ll}
\rd s^{2} = \frac{1}{\hat{c}(2-\hat{c}f)}\left(-\rd t^{2} + (\rd x^{1})^{2} \right) + (\rd x^{2})^{2} + \cdots + (\rd x^{9})^{2}\,,
\\
B = -\frac{1 - \hat{c}f}{\hat{c}(2-\hat{c}f)}\,\rd t\wedge\rd x^{1}\,,
\\
e^{-2\phi} = e^{-2\phi_0} \hat{c}(2-\hat{c}f)\,.
\ea\label{dualgeommetric}
\ee
For non-vanishing $\hat{c}$ this metric is non-degenerate and well-defined. 
In the geometric representation \eqref{dualgeommetric} the limit $\hat{c}\rightarrow 0$ appears inconsistent, as the fields in \eqref{dualgeommetric} become either singular or vanish everywhere. Nevertheless, the generalized metric \eqref{dualgeom} is well-defined even in the limit $\hat{c}\rightarrow 0$, and becomes
\be
\cH_{MN} =\left(\ba{cccc}
0 & 0 & \cE^{\alpha}{}_{\beta} & 0 \\
0 & \delta^{ij} & 0 & 0 \\
-\cE_{\alpha}{}^{\beta} & 0 & f\eta_{\alpha\beta} & 0 \\
0 & 0 & 0 & \delta_{ij} \ea\right)\,.
\label{dualgeom2}
\ee

The sigma model~(\ref{Lagrangian}) then takes the following form: 
\be
\textstyle{\frac{1}{4\pi\alpha^{\prime}}}\cL={\textstyle{\frac{1}{2\pi\alpha^{\prime}}}}\cL^{\prime\prime}\,,
\label{tensionLprimeprime}
\ee
\be
\ba{rl}
\cL^{\prime\prime}=&-\quarter\sqrt{-h}h^{ab}
\partial_{a}X^{\alpha}\partial_{b}X^{\beta}\eta_{\alpha\beta}f(X)-\half\sqrt{-h}h^{ab}
\partial_{a}X^{i}\partial_{b}X_{i}+\half\epsilon^{ab}\partial_{a}\tX_{\mu}\partial_{b}X^{\mu}
\\
{}&+\half\sqrt{-h}(A_{a\alpha}-\partial_{a}\tilde{X}_{\alpha})\left(
\cE^{\alpha}{}_{\beta}h^{ab}\partial_{b}X^{\beta}+\textstyle{\frac{1}{\sqrt{-h}}}\epsilon^{ab}\partial_{b}X^{\alpha}\right)\\
{}&-\quarter\sqrt{-h}h^{ab}\left(\partial_{a}\tX_{i}+\textstyle{\frac{1}{\sqrt{-h}}}\epsilon_{a}{}^{c}\partial_{c}X_{i}-A_{ai}\right)\left(\partial_{b}\tX^{i}+\textstyle{\frac{1}{\sqrt{-h}}}\epsilon_{b}{}^{e}\partial_{e}X^{i}-A_{b}{}^{i}\right)\,.
\ea
\label{Lprimeprime}
\ee
Note that while the $\SO(8)$ sector of $\{X^{i}, \tilde{X}_{i}, A_{ai}\}$ agrees%
, up to constraints, with the standard sigma model, 
the $\SO(1,1)$ sector of $\{X^{\alpha}, \tilde{X}_{\alpha}, A_{a\alpha}\}$ takes a novel, more exotic form.  In particular, the gauge field appears quadratically in \eqref{Lprimeprime} for the non-degenerate $\SO(8)$ sector, but only linearly in the $\SO(1,1)$ sector.

Integrating out all the gauge fields, the doubled yet gauged  sigma model reduces to
\be
\textstyle{\frac{1}{2\pi\alpha^{\prime}}}\left[-\quarter\sqrt{-h}h^{ab}
\partial_{a}X^{\alpha}\partial_{b}X^{\beta}\eta_{\alpha\beta}f(X)
-\half\sqrt{-h}h^{ab}
\partial_{a}X^{a}\partial_{b}X_{a}+\half\epsilon^{ab}\partial_{a}\tX_{\mu}\partial_{b}X^{\mu}\right]\,,
\label{Ldeg}
\ee
where now the two of the `ordinary' coordinate fields are constrained to satisfy a self-duality  constraint,
\be
\partial_{a}X^{\alpha}+\textstyle{\frac{1}{\sqrt{-h}}}\epsilon_{a}{}^{b}\cE^{\alpha}{}_{\beta}\partial_{b}X^{\beta}=0\,.
\label{SDcE}
\ee
This is in contrast to  the  non-degenerate $\SO(8)$ sector of which the  ordinary and the dual coordinate fields are related by the standard self-duality relation~\req{eq:self-duality}. Note that upon the self-duality~(\ref{SDcE}), the second line of (\ref{Lprimeprime}) vanishes. 

To summarize, even for the degenerate sector for which the Riemannian metric is ill-defined, there exists a sigma model type Lagrangian description involving a self-duality constraint.

\subsection{The Gomis-Ooguri Background in Double Field Theory}
The doubled geometry \req{dualgeom2} is in fact a solution for \emph{any} harmonic function $f$ of the variables $x^i$.
The simplest case is to let $f$ be a constant, and we show here that the doubled sigma model on this background is in fact precisely the Gomis-Ooguri non-relativistic string.

This background is flat, and thus exists for the bosonic string as well, so we return to the general case of $D$ spacetime dimensions.
The sigma model coordinates split into two types, $X^\alpha = (X^0,X^1)$ and $X^i$.
The $X^i$ coordinates form $(D-2)$-dimensional Euclidean space as usual, so we focus on $X^\alpha$.
Denoting the generalized coordinate for $X^\alpha$ by $X^A = (\tilde X_\alpha,X^\alpha)$, the background $g_{\alpha\beta}=G\,\eta_{\alpha\beta}$, $B_{\alpha\beta}=(G-\mu)\,\epsilon_{\alpha\beta}$ considered in section~\ref{ssec:NRCS} corresponds to the doubled geometry
\begin{equation}
\cH_{AB} = \left(\ba{cc} \frac{1}{G}\eta^{\alpha\beta} & \frac{G-\mu}{G}\cE^\alpha_{\phantom\alpha\beta} \\
-\frac{G-\mu}{G}\cE_{\alpha}^{\phantom\alpha\beta} & 2\mu\,\eta_{\alpha\beta} \ea\right) \,.
\end{equation}
Taking the limit $G\to\infty$ and identifying $f=2\mu$, we obtain precisely the flat non-Riemannian background~\req{dualgeom2}.
Looking at the DFT dilaton field $d$, we also have in our coordinate system
\begin{equation}
e^{-2d} = \sqrt{g}e^{-2\phi} = G\, g_s^{-2} = g_0^{-2}\,,
\end{equation}
so that $d$ remains finite in the $G\to\infty$ limit.
Thus the Gomis-Ooguri limit is a non-singular configuration of DFT, and as such we expect that the worldsheet DFT sigma model should provide a manifestly non-singular sigma model description on this backgrounds.

In fact, it is now straightforward to see that, after gauge-fixing, the gauge fields $A_{a\alpha}$ of the previous section are nothing other than the $\beta,\bar\beta$ Lagrange multipliers of the Gomis-Ooguri sigma model, and that the modified self-duality constraint \req{SDcE} is the constraint imposed by them.

Let us make this more apparent.
We now work in conformal gauge on the worldsheet, $h_{ab}=e^{2\phi}\eta_{ab}$, and switch to light-cone coordinates, both on the worldsheet ($\sigma^\pm=\sigma^1\pm\sigma^0$) and in the target spacetime ($\gamma = X^1 + X^0$, $\bar\gamma = X^1 - X^0$).
Next, recall that we have a gauge symmetry $\delta\tilde X_\alpha = \tilde\lambda_\alpha$, $\delta A_{a\alpha} = \p_a\tilde\lambda_\alpha$, which we can fix completely (including large gauge transformations) by imposing $\tilde X_\alpha = 0$.
Inserting these conditions into the $(\tilde X_\alpha, X^\alpha,A_{a\alpha})$ sector of the Lagrangian~\req{Lagrangian}, we obtain
\be
\cL_{0,1} = -2\p_-\gamma A_+ + 2\p_+\bar\gamma\bar A_- - f(\p_+\gamma\p_-\bar\gamma + \p_+\bar\gamma\p_-\gamma) \,.
\ee
Make the identifications $A_+=\beta$, $A_-=-\bar\beta$, and $f=2\mu$, and perform a Wick rotation on the worldsheet coordinates, $(\sigma^+,\sigma^-)\rightarrow (z,\bar z)$.
Then, after an integration by parts that (having vanishing boundary contributions in our background) is harmless, we obtain precisely the action of the $(\beta,\gamma,\bar\beta,\bar\gamma)$ sector of the Gomis-Ooguri string found in equation~\req{eq:GO-action}.

The construction in section~\ref{ssec:GalileanSigmaModel} of a similar GO-like background using T-duality raises an important question. In standard toroidal compactifications, T-duality has the effect of exchanging winding number $w$ and momentum $n$ on the duality circle.
In the GO dispersion relation \req{eq:GO-spectrum}, however, they appear asymmetrically.
Indeed, $n$ does not appear at all;
instead, it contributes only to the level-matching condition $wn=N-\tilde N$.
Nonetheless, because the background \req{dualgeom2} is non-degenerate in Double Field Theory, $O(D,D)$ invariance still guarantees us a T-dual theory (\textit{c.f.~}foonote \ref{footR}), which is equivalent when stringy excitations are included.
We can now see the correct interpretation of this duality: in the $c\to\infty$ limit of equation \req{eq:421}, T-duality becomes an identity relating two worldsheet theories with different field content --- a light-cone sigma model, and the GO sigma model.

\subsection{Generalized Diffeomorphisms and Galilean Symmetries%
\label{sec:galilean-symmetry}}
Having established that the Gomis-Ooguri string arises as a consistent background of both the bulk action and the worldsheet sigma model of Double Field Theory, we now turn to the question of symmetries.
The algebra of infinitesimal symmetries of the spectrum of fluctuations around some background geometry is typically determined by the algebra of Killing vector fields of that geometry, \emph{i.e.} the gauge parameters $\xi^M$ for which the generalized Lie derivatives $\hat\cL_\xi\cH_{AB}$ and $\hat\cL_\xi d$ both vanish.

As the Gomis-Ooguri string has a Galilean-invariant spectrum, it is natural to expect that the Galilean algebra (or rather its central extension, the Bargmann algebra) is realized by the generalized Killing fields of the Gomis-Ooguri background in DFT.
The Bargmann algebra has generators $H$, $P_i$, $M_{ij}$, $B_i$, $N$ corresponding respectively to time and space translations, rotations, Galilean boosts, and the particle number.
The non-vanishing commutators take the form
\begin{equation}
\begin{array}{c}
[B_i,H] = P_i \qquad
[B_i,P_j] = \delta_{ij}N  \qquad
[M_{ij},P_k] = \delta_{ik}P_j - \delta_{jk}P_i \\
{}[M_{ij},B_k] = \delta_{ik}B_j - \delta_{jk}B_i \qquad
[M_{ij},M_{k\ell}] = \delta_{ik}M_{j\ell} - \delta_{i\ell}M_{kj} - \delta_{jk}M_{i\ell} + \delta_{j\ell}M_{ik} \: .
\end{array}
\label{bargmann}
\end{equation}

A natural representation of vector fields $\xi^M=(\tilde\lambda_\mu,\lambda^\mu)$ in doubled geometry is to write $\xi = \xi^M\p_M = \lambda^\mu\p_\mu + \tilde\lambda_\mu\tilde\p^\mu$.
Imposing the Killing condition
\begin{equation}
\hat \cL_\xi\cH_{AB} = \xi^C\p_C\cH_{AB} + \cH_{AC}(\p_B\xi^C-\p^C\xi_B) + \cH_{CB}(\p_A\xi^C - \p^C\xi_A) = 0
\end{equation}
on the Gomis-Ooguri background, we find the following gauge parameters generating global symmetries:
\begin{align}
H &= -\p_t & Q &= -\p_1 \\
P_i &= -\p_i & N &= -\tilde\p^1 \\
M_{ij} &= -(x^i\p_j - x^j\p_i) & B_i &= -t\p_i - x^i\tilde\p^1 \,,
\end{align}
together with pure $B$-field gauge transformations acting trivially on physical states.
Unlike in standard geometry, DFT gauge transformations are commuted using the C-bracket
\be
[\xi,\eta]_\text{C} = \xi^A(\p_A\eta^B)\p_B - \eta^A(\p_A\xi^B)\p_B 
- \frac{1}{2}\xi_A(\p^B\eta^A)\p_B + \frac{1}{2}\eta_A(\p^B\xi^A)\p_B \,,
\ee
and it is straightforward to verify that under the C-bracket these transformations close on the Bargmann algebra \req{bargmann} (supplemented by the $U(1)$ generator $Q$), giving a natural realization of Galilean symmetry within the doubled geometry of Double Field Theory.

\subsection{A Doubled Geometry with Schr\"odinger Conformal Symmetry}\label{ssec:Schroedinger}
Let us now take a short detour from our main line of development, and consider potential applications of this new realization of Galilean symmetry in gravity.
Likely the most important manifestation of non-relativistic symmetry algebras within gravity and string theory during the past several years lies in the attempt to construct gravitational duals to non-relativistic field theories.
Just as many of the important properties of the original AdS/CFT proposal relied heavily on the relationship between the geometric symmetries of a gravitational background and the global symmetries of the dual theory, the strategy here was to find geometries whose geometric symmetries reproduce the algebra of symmetries of a dual non-relativistic scale-invariant theory.
The most restrictive -- and, thus, most useful -- symmetry algebras are those whose symmetry algebras contain the Bargmann algebra \req{bargmann}, and it was precisely theories with these symmetries which initiated the study of the non-relativistic gauge/gravity correspondence in~\cite{Son:2008ye,Balasubramanian:2008dm}.
We restrict ourselves to this case in the remainder.

Unlike relativistic theories at scale-invariant fixed points, there is no reason for time and space to scale the same way in a non-relativistic field theory.
The parameter characterizing this discrepancy is the \emph{dynamical critical exponent} $z$, and the scaling symmetry of the dual field theory takes the form
\begin{equation}
t \mapsto \lambda^z \, t\,, \qquad\qquad \vec x \mapsto \lambda\, \vec x \; .
\end{equation}
The infinitesimal generator $D$ of these transformations has the following commutators (C-brackets) with the Bargmann generators:
\begin{equation}
[D,H] = zH\,,
\qquad
[D,P_i] = P_i\,,
\qquad
[D,B_i] = -(z-1)B_i\,,
\qquad 
[D,N] = -(2-z)N\,,
\qquad
[D,M_{ij}] = 0
\; . 
\label{eq:scale-invariant-Bargmann}
\end{equation}
Finally, for the special case $z=2$ an additional generator $C$ with commutators
\begin{equation}
[C,H] = D\,, \qquad
[C,D] = C\,, \qquad
[C,P_i] = B_i\,, \qquad
[C,B_i] = [C,M_{ij}] = [C,N] = 0\,, 
\end{equation}
can be added to the algebra.
The algebra spanned by $(H,D,C,P_i,B_i,M_{ij},N)$ is called the \emph{Schr\"odinger conformal algebra}, and $C$, the special conformal generator.

Section~\ref{sec:galilean-symmetry} gave a natural realization of the Galilean algebra in DFT, and it is natural to expect that one can also write down non-Riemannian DFT geometries whose symmetries are the Schr\"odinger conformal algebra, as we will now show. 
Here, we assume the section condition $\tilde\p^\mu = 0$, and as before, we split the generalized coordinates as $M=(A,I)$, although we will write all expressions in terms of the original spacetime coordinates $x^\alpha=(t,x^1)$ rather than  light-cone coordinates.
Further breaking up the coordinates $x^i=(x^m,u)$, consider a spacetime configuration given by
\begin{equation}
\ba{c}
\cH_{AB} = \left(\ba{cc} 0 & \sigma^\alpha_{\phantom\alpha\beta}(u) \\ \sigma_\alpha^{\phantom\alpha\beta}(u) & \cH_{\alpha\beta} \ea\right)\,,
\qquad
\cH_{IJ} = \left(\ba{cc} u^2\,\delta^{ij} & 0 \\ 0 & u^{-2}\,\delta_{ij} \ea\right)\,,
\qquad
\cH_{AI} = 0\,, 
\label{schroedinger-geometry}
\\
\cH_{\alpha\beta} = \left(\ba{cc} -\frac{1}{u^{2z}} & 0 \\ 0 & u^{4-2z} \ea\right)\,,
\qquad
\sigma^\alpha_{\phantom\alpha\beta}(u) = (\sigma_\beta^{\phantom\beta\alpha}(u))^T = \left(\ba{cc} 0 & -u^2 \\ -\frac{1}{u^2} & 0 \ea\right) 
\; .
\ea
\end{equation}
Here $u$ plays the role of the radial coordinate in the standard Poincar\'e patch.
The action of a DFT gauge transformation $\xi$ on $\cH$ takes the form
\begin{align}
\hcL_\xi\cH_{\alpha\beta} &= 
2\sigma^\gamma_{(\alpha}\p_{\beta)}\tilde\lambda_\gamma - 2\p_\gamma\tilde\lambda_{(\alpha}\sigma_{\beta)}^\gamma + 2\cH_{\gamma(\alpha}\p_{\beta)}\lambda^\gamma + \lambda^u\p_u \cH_{\alpha\beta}
\,, \span\span \hspace{\stretch{1}} \hcL_{\xi}\cH^{\alpha\beta} = \, 0 \,,
\nonumber\\
\hcL_{\xi}\cH_{\alpha}^{\phantom\alpha\beta} &= \p_\alpha\lambda^\gamma\sigma_{\gamma}^\beta - \sigma_{\alpha}^\gamma\p_\gamma\lambda^\beta + \lambda^u\p_u\sigma_\alpha^\beta\,, &
\hcL_{\xi}\cH_{\alpha}^{\phantom{\alpha}i} &= \cH^{ij}(\p_\alpha\tilde\lambda_j - \p_j\tilde\lambda_\alpha) - \sigma_\alpha^\gamma\p_\gamma\lambda^i \,, \\
\hcL_{\xi}\cH_{ij} &= 2\cH_{k(i}\p_{j)}\lambda^k + \lambda^u\p_u\cH_{ij}\,, &
\hcL_{\xi}\cH_i^{\phantom{i}j} &= \cH^{jk}(\p_i\tilde\lambda_k - \p_k\tilde\lambda_i) 
\nonumber\\
\hcL_{\xi}\cH_{\alpha i} &= \cH_{ij}\p_\alpha\lambda^j + \cH_{\alpha\gamma}\p_i\lambda^\gamma + \sigma_\alpha^\gamma(\p_i\tilde\lambda_\gamma - \p_\gamma\tilde\lambda_i) &
\hcL_{\xi}\cH^{\alpha i} &= -\cH^{ij}\p_j\lambda^\alpha
\end{align}
and those related to them by the constraint $\cH_{A}^{\phantom{A}C}\cH_{B}^{\phantom{B}{D}}\cD_{CD} = \cJ_{AB}$.
Solving $\hcL_\xi\cH = 0$ for $\xi$, we find the following generalized Killing vectors:
\begin{align}
H &= - \p_t\,, &
D &= -zt\p_t - x^m\p_m - u\p_u - (z-2)x^1\p_1 \,, \nonumber\\
P_m &= - \p_m\,,  &
B_m &= -t\p_m - x^m\tilde\p^1\,, \\
N &= -\tilde\p^1\,, &
M_{mn} &= -(x^m\p_n - x^n\p_m) \,. \nonumber
\end{align}
(plus $\p_1$ and the trivial $B$-field gauge transformations), comprising the scale invariant Galilean symmetry algebra with dynamical critical exponent $z$.
Caution must be taken when interpreting the C-bracket commutators of these generators.
For example, the C-bracket of $D$ and $B_m$ takes the form
\be
[D,B_m]_\text{C} = -(z-1)B_m - \frac{z-2}{2}(x^m\tilde\p^1 + x^1\tilde\p^m) =-(z-1)B_m - \frac{z-2}{2}\partial^{M}(x^1x^m)\partial_{M}\,.
\ee
Here it is important to note that the potentially  ``anomalous" term, $\p^M(x^1x^m)\p_M$, corresponds to the kernel of the generalized Lie derivative, \textit{i.e.~}$\hcL_{\partial^{M}(x^1x^m)}=0$, and is also a trivial coordinate gauge symmetry.
Thus it can be set equal to zero when acting on physical states, and with this in mind the commutators take the desired form \req{eq:scale-invariant-Bargmann}.

When $z=2$ there is an additional symmetry generator,
\begin{align}
C &= -t^2\p_t - tx^m\p_m - tu\,\p_u - \frac{1}{2}(x^2 + u^2)\tilde\p^1 .
\end{align}
It is straightforward to check that in this case, under the C-bracket the generators satisfy the full Schr\"odinger conformal algebra.

As in the case of the Gomis-Ooguri background, this geometry is related by T-duality to the well-known geometric Schr\"odinger background~\cite{Son:2008ye}, and so in this sense \req{schroedinger-geometry} is not a fundamentally new configuration within string theory.
Nonetheless, within the realm of DFT -- where the choice of section condition is part of the definition of the theory -- this background provides a novel (non-)geometric realization of the Schr\"odinger algebra in terms of the generalized diffeomorphisms of DFT.

\newpage

\section{Fluctuations in the Non-Relativistic Closed String Background\label{sec:4}}
%
In the previous section we saw that the structures and symmetries of DFT were capable of incorporating the Gomis-Ooguri limit of the closed string: we showed that the DFT sigma model reproduces the Gomis-Ooguri worldsheet theory, and gave an embedding of the Gomis-Ooguri limit in DFT.
We further showed that Galilean invariance embeds naturally into the symmetries of DFT, and further that there exist doubled geometries whose generalized isometries realize the Schr\"odinger conformal algebra. 

In this section we turn to dynamics.
We show in particular that the fluctuations equations of DFT correctly reproduce the portion of the string spectrum with trivial (massless sector) oscillator excitations, i.e., the part of the string spectrum which descents from the massless sector of the relativistic (super)string.

\subsection{Spectrum of Double Field Theory on the Gomis-Ooguri Background\label{ssec:GOspectrum}}
DFT is the effective field theory of the massless modes of the string in a particular background, depending on the section condition, and provided that all scales in the corresponding geometry are much larger than string scale. 
Because the non-relativistic limit taken in section \ref{ssec:NRCS} sends the radius of the compact direction $x^1$ to infinity (by sending the speed of light $c$ to infinity, c.f. the explanation around eq.~\ref{eq:421}), the natural section condition to describe the Kaluza-Klein sector of the Gomis-Ooguri string is $\tilde\p^\mu = 0$, which projects out the winding modes. This is the primary section condition we are going to consider in this subsection. 
We will consider the alternative T-dual section condition when we analyze the spectrum of winding modes in the T-dual frame in section \ref{ssec:Tdualspec}.

We start by adopting light-cone coordinates $x^\pm = \frac{1}{\sqrt 2}(t \pm x^1)$ and their dual coordinates $\tilde x^\pm = \frac{1}{\sqrt 2}(\tilde t \pm \tilde x^1)$.
In the $(x^+,x^-)$ basis, the tensors in \req{Lprimeprime} take the matrix form
\begin{equation}
(\cE^\alpha_{\phantom\alpha\beta}) = \left(\ba{cc} -1 & 0 \\ 0 & 1 \ea \right)\,,
\qquad
\eta_{\alpha\beta} = \left(\ba{cc} 0 & -1 \\ -1 & 0 \ea\right) \,.
\end{equation}
%
We also introduce the alternative notation 
$\sigma^\alpha_\beta = \cE^\alpha_{\phantom\alpha\beta}$, with indices unordered;
in this background there is no background metric with which to raise or lower indices,
so this causes no ambiguity.
For clarity we consider an arbitrary constant metric $g_{ij}$ in the $x^i$ directions, so that the background is
\begin{equation}
\cH_{MN} =\left(\ba{cccc}
0 & 0 & \cE^{\alpha}{}_{\beta} & 0 \\
0 & g^{ij} & 0 & 0 \\
-\cE_{\alpha}{}^{\beta} & 0 & f\eta_{\alpha\beta} & 0 \\
0 & 0 & 0 & g_{ij} \ea\right)\,.
\label{eq:go-geom}
\end{equation}
We wish to make the connection to the Gomis-Ooguri string, and so for simplicity we assume that $f$ is constant, although the more general case can be treated with similar methods.

Fluctuations are solutions to the linearized equations of motion \req{EOMf1} and \req{EOMf2}, obtained by replacing $\cH_{AB} \mapsto \cH_{AB} + h_{AB}$, $d \mapsto d + \psi$ and expanding to first order in the perturbations $h_{AB}$ and $\psi$.
The $\ODD$ condition $\cH_A^{\phantom AC}\cH_{CB}=\cJ_{AB}$ imposes the constraint \req{deltaPconstraint} on perturbations.
In terms of $h_{AB}$ this is
\begin{equation}
h_{AB} = -\cH_A^{\phantom AC}h_{CD}\cH^D_{\phantom DB} \,,
\end{equation}
which in the chosen basis takes the explicit form 
\begin{align}
h^{\alpha\beta} &= -\sigma^\alpha_\gamma h^{\gamma\delta} \sigma^\beta_\delta\,, &
h^\alpha_{\phantom \alpha\beta} &= -\sigma^\alpha_\gamma h^\gamma_{\phantom\gamma\delta}\sigma^\delta_\beta - f \sigma^\alpha_\gamma h^{\gamma\delta}\eta_{\delta\beta} \,,\\
h^{\alpha i} &= -\sigma^\alpha_\beta h^\beta_{\phantom\beta j}g^{ji}\,, &
h_{\alpha\beta} &= -\sigma_\alpha^\gamma h_{\gamma\delta}\sigma_\beta^\delta - f\sigma_\alpha^\gamma h_\gamma^{\phantom \gamma\delta}\eta_{\delta\beta} - f\eta_{\alpha\gamma}h^\gamma_{\phantom \gamma\delta}\sigma^\delta_\beta - f^2\eta_{\alpha\gamma}h^{\gamma\delta}\eta_{\delta\beta} \,,
\nonumber\\
h_i^{\phantom ij} &= b_{ik}g^{kj}\,, &
h_{\alpha i} &= -f\eta_{\alpha\gamma}h^{\gamma j}g_{ji} - \sigma_\alpha^\beta h_{\beta}{}^j g_{ji}\,, \\
h^i_{\phantom{i}j} &= -g^{ik}b_{kj}\,, &
h^{ij} &= -g^{im}h_{mn}g^{nj}\,, 
\nonumber
\end{align}
with $b_{ij}$ an antisymmetric tensor.

The first equation 
implies we can write $h^{\alpha\beta} = \hat h\,\eta^{\alpha\beta}$, so that $h^{++}=h^{--}=0$ and $h^{+-}=-\hat h$.
It is possible to solve the rest of these constraints explicitly, but it is simpler to begin by fixing the gauge.

\subsubsection{Linearized Gauge Symmetries}
Assuming that $f$ is constant, after linearization the generalized diffeomorphisms take the form
\begin{equation}
\delta h_{AB} = \hat\cL_\xi\cH_{AB} = \cH_{AC}\p_B\xi^C + \cH_{CB}\p_A\xi^C - \cH_{AC}\p^C\xi_B - \cH_{CB}\p^C\xi_A .
\end{equation}
In our decomposition, with $\xi^M = (\tilde\lambda_\mu,\lambda^\mu)$, 
\begin{align}
\delta h_{\alpha\beta} &= 2\sigma^\gamma_{(\alpha}\p_{\beta)}\tilde\lambda_\gamma - 2\p_\gamma\tilde\lambda_{(\alpha}\sigma_{\beta)}^\gamma + 2f\eta_{\gamma(\alpha}\p_{\beta)}\lambda^\gamma\,, &
\delta h^{\alpha\beta} &= 0 
\label{delta-hab}\,, \\
\delta h_{\alpha}^{\phantom\alpha\beta} &= \p_\alpha\lambda^\gamma\sigma_\gamma^\beta - \sigma_\alpha^\gamma\p_\gamma\lambda^\beta\,, &
\delta h^{\alpha i} &= -g^{ij}\p_j\lambda^\alpha\,, \\
\delta h_\alpha^{\phantom\alpha i} &= g^{ij}(\p_\alpha\tilde\lambda_j - \p_j\tilde\lambda_\alpha) - \sigma_\alpha^\gamma\p_\gamma\lambda^i\,, &
\delta h^\alpha_{\phantom\alpha i} &= \sigma^\alpha_\gamma\p_i\lambda^\gamma\,, \\
\delta h_{\alpha i} &= g_{ij}\p_\alpha\lambda^j + f\eta_{\alpha\gamma}\p_i\lambda^\gamma + \sigma_\alpha{}^\beta(\p_i\tilde\lambda_\beta - \p_\beta\tilde\lambda_i) \,. &
\end{align}
These allow us to choose the following gauge-fixing conditions.
(There are residual symmetries we use later.)
\begin{enumerate}
\item Use $\lambda^\alpha$ to fix $h^\alpha_{\phantom\alpha\beta} = -\frac{1}{2}f\hat h\sigma^\alpha_\beta$
\item Use $\tilde\lambda_\alpha$ to fix $h_{\alpha\beta} = 0$.
\item Use $\tilde\lambda_i$ and $\lambda^i$ to fix $h_{\alpha i}=0$.
\end{enumerate}
Imposing these conditions leaves us with the independent variables
$\hat h, h_{ij}, b_{ij}, h^\alpha_{\phantom\alpha i}$,
%
%
together with the relations 
\begin{align}
h^\alpha_{\phantom\alpha\beta} &= -\frac{1}{2}f\,\hat h\,\sigma^\alpha_\beta\,, &
h^{\alpha i} &= -\sigma^\alpha_\gamma h^\gamma_{\phantom\gamma j}g^{ji}\,, &
h_\alpha^{\phantom\alpha i} &= -f\eta_{\alpha\gamma}h^\gamma_{\phantom\gamma j}g^{ji} .
\label{gauge-fixing}
\end{align}

\subsubsection{Gauge-Fixing and the Linearized Equations of Motion}
As our background is flat, $\nabla_A=\p_A$.
In terms of $h_{AB}$ and $\psi$, the linearized equations of motion (\ref{EOMf1},\ref{EOMf2}) thus take the form
\be
\p^{A}\p^{B}h_{AB} - 4\cH^{AB}\p_{A}\partial_{B}\psi = 0\,,
\label{EOMh1}
\ee
\be
(P_{A}^{\phantom AC}\bar\Delta_{B}^{\phantom BD} - \Delta_{A}^{\phantom AC}\brP_{B}^{\phantom BD})h_{CD} + 8P_A^{\phantom AC}\bar P_B^{\phantom BD}\p_C\p_D\psi = 0\,,
\label{EOMh2}
\ee
where the differential operators \eqref{DeltabrDelta} become 
\begin{align}
\Delta_B^{\phantom BD} &= (P_B^{\phantom BD}P^{EF} - 2P_B^{\phantom BE}P^{DF})\p_E\p_F\,, \\
\bar\Delta_B^{\phantom BD} &= (\bar P_B^{\phantom BD}\bar P^{EF} - 2\bar P_B^{\phantom BE}\bar P^{DF})\p_E\p_F \,.
\end{align}
With this form of the differential operators, the fluctuation equations \eqref{EOMh1}, \eqref{EOMh2} explicitly read
\begin{equation}
\begin{split}
\cE_\psi &= \p^A \p^B h_{AB} - 4\cH^{AB}\p_A \p_B \psi = 0\,, \\
\cE_{AB} &= \bigg[ 2(P_A^C\bar P_B^E\bar P^{DF} - \bar P_B^D P_A^E P^{CF}) + P_A^C \bar P_B^D \cH^{EF} \bigg] \p_E\p_F h_{CD} - 8P_A^C\bar P_B^D \p_C \p_D \psi = 0 .
\end{split}
\end{equation}

We now impose the gauge-fixing condition above, and expand in plane waves of momentum $p_\mu=(p_+,p_-,k_i)$, 
setting $h_{AB}(x)=h_{AB}e^{ip_+x^+ + ip_-x^- + ik_ix^i}$.
We further decompose the fluctuations into spatially transverse and longitudinal components 
\begin{align}
h_{ij} &= h^\perp_{ij} + k_i\zeta^\perp_j + k_j\zeta^\perp_i + (k_ik_j - \frac{1}{D-2}k^2g_{ij})\rho + \frac{1}{D-2}hg_{ij}\,, \\
b_{ij} &= b^\perp_{ij} + k_i\chi^\perp_j - k_j\chi^\perp_i\,, \\
h_i^{\phantom i \alpha} &= h_i^{\perp\alpha} + k_i\phi^\alpha\,, 
\end{align}
satisfying the transversality constraints $k^ih^\perp_{ij} = k^ib^\perp_{ij} = k^i \zeta^\perp_i = k^i h_i^{\perp\alpha} =0$, 
where we raised the index on the momentum by $k^i = g^{ij}k_j$. 
Using this decomposition, the complete equations of motion are expressed in components in the following way:
\begin{equation}
\cE_\psi = 2p_+p_-\hat h + 2k^2(p_-\phi^- - p_+\phi^+) + \frac{1}{D-2}k^2[h - (D-3)k^2\rho + 4(D-2)\psi]\,, \label{eompsi} 
\end{equation}
\begin{align}
\cE^{-+} &= k^2\hat h\,, & 
\cE^-_{\phantom -+} &= 2p_+(k^2\phi^- + p_+\hat h) \label{eom-+1}\,, \\
\cE_{-+} &= fk^2(p_-\phi^- - p_+\phi^+ + \frac{1}{4}f\hat h) + 8p_+p_-\psi\,, &
\cE_-^{\phantom -+} &= 2p_-(k^2\phi^+ - p_-\hat h) \label{eom-+2}\,, \\
\cE_{-i} &= p_-k^m(h_{mi}-b_{mi}) + 2p_-^2h_i^- + \frac{f}{2}k^2h_i^{\perp+} + 4p_-k_i\psi\,, &
\cE^-_{\phantom -i} &= -k^2 h_i^{\perp -} + p_+k_i\hat h \label{eom-i}\,, \\
\cE_{i+} &= p_+k^m(h_{mi}+b_{mi}) - 2p_+^2h_i^+ - \frac{f}{2}k^2h_i^{\perp-} + 4p_+k_i\psi\,, &
\cE_i^{\phantom i+} &= -k^2 h_i^{\perp +} - p_-k_i\hat h \label{eomi+}\,,
\end{align}
\begin{multline}
\qquad 
\cE_{ij} = \frac{1}{2}k_i \left[ 2p_-h_j^- + g^{mn}k_m(h_{nj} - b_{nj}) \right] \\
 - \frac{1}{2}k_j\left[ 2p_+h_i^+ - g^{mn}k_m(h_{ni} + b_{ni}) \right] 
- k^2(h_{ij}-b_{ij}) + 2k_ik_j\psi . \qquad
\label{eomij}
\end{multline}

\subsubsection{Solution of the Linearized Equations of Motion}
To build normalizable wave packets, there must be non-vanishing solutions to these equations with $k^2\ne 0$.
Moreover, to be propagating at least one of $p_-$ and $p_+$ must be non-zero.
Assuming these conditions, we now prove that all components of $h_{AB}$ vanish.

Equation \req{eom-+1} immediately implies that $\hat h=0$.
Supplemented by the gauge conditions \req{gauge-fixing}, we find that
\begin{equation}
h_{\alpha\beta}=h^\alpha_{\phantom\alpha\beta}=h^{\alpha\beta}=0\, . 
\end{equation}
Equations (\ref{eom-+1},\ref{eom-+2}) now become $p_+\phi^-=p_-\phi^+=0$.
We can use this, by multiplying $\cE_{-+}$ by $p_+p_-$, to conclude that $p_-^2p_+^2\psi = 0$.
Hence $p_-p_+\psi$, must vanish.

The vanishing of $\cE_{-+}$ reduces to $fk^2(p_-\phi^--p_+\phi^+)=0$.
Now there are two cases:
If $f\ne 0$, we multiply $\cE_{-+}$ by $p_+$ and find that $p_+\phi^+=0$; similarly, $p_-\phi^-=0$.
Thus we have $p_+\phi^+=p_+\phi^-=p_-\phi^+=p_-\phi^-=0$; but by assumption at least one of $p_+$ and $p_-$ is non-zero, so we conclude that $\phi^+=\phi^-=0$.
If, on the other hand, $f=0$, then we consider the gauge transformations $\lambda^\pm$.
This gauge symmetry was fixed by imposing conditions on $h^\alpha_\beta$ and $h_{\alpha\beta}$.
From \req{delta-hab} we see that if $f=0$, $\lambda$ is a residual gauge symmetry whenever $p_+\lambda^-=p_-\lambda^+=0$.
Setting $\lambda^+ = \phi^+$ and $\lambda^- = -\phi^-$, the equations of motion imply that indeed $\lambda^\pm$ are residual gauge transformations.
Under these transformations, $\delta\phi^{\pm} = -\phi^\pm$.
Thus we may always take $\phi^\pm = 0$.
The right-hand equations in (\ref{eom-i},\ref{eomi+}) are now $h_i^\perp=0$, so we conclude that 
\begin{equation}
h_i^\alpha = h_{\alpha i} = h^{\alpha i} = 0.
\end{equation}

Now \req{eompsi} reduces to 
\begin{equation}
h+(D-3)k^2\rho + 4(D-2)\psi = 0.
\end{equation}
Using this, the left-hand sides of (\ref{eom-i},\ref{eomi+}) reduce to $p_-(\zeta_i^\perp - \chi_i^\perp) = p_+(\zeta_i^\perp + \chi_i^\perp) = 0$.
The condition used to gauge-fix $\lambda^i$ and $\tilde\lambda_i$ was $h_{\alpha i}=0$, which transforms as 
$\delta h_{\pm i} = p_\pm(g_{ij}\lambda^j\pm \tilde\lambda_i)$.
If we set $\lambda^i = -g^{ij}\zeta_j^\perp$, $\tilde\lambda_i = - \chi_i^\perp$, the equations of motion imply that
these are residual gauge transformations.
For transverse gauge parameters, $\delta\zeta_i^\perp = g_{ij}\lambda^j$ and $\delta\chi_i^\perp = \tilde\lambda_i$, so applying the transformation fixes $\zeta_i^\perp = \chi_i^\perp = 0$.

The final equation \req{eomij} now reduces to 
\begin{equation}
-k^2 h_{ij}^\perp - k^2 b_{ij}^\perp + k_ik_j \Bigl[ 2\psi (D-2)h - k^2\rho + 2\psi \Bigr] - \frac{1}{D-2}k^2(h - k^2\rho) = 0 .
\end{equation}
The vanishing of the trace implies $\frac{D-3}{D-2}(k^2\rho - h) + 2\psi = 0$, while the trace-free scalar part gives
$h - k^2\rho + 2(D-2)\psi = 0$.
These two equations, together with $\cE_\psi=0$, imply that $h = \rho = \psi = 0$.

Finally, transverse part of \req{eomij} gives $h_{ij}^\perp = b_{ij}^\perp = 0$.
Thus we conclude that
\begin{equation}
h_{ij} = b_{ij} = 0 .
\end{equation}
This shows that there are no normalizable fluctuations around the Gomis-Ooguri background satisfying the section condition $\tilde\p^\mu = 0$, in agreement with the dispersion relation \req{eq:GO-spectrum}.

\subsection{The Spectrum in the T-dual Frame}\label{ssec:Tdualspec}
In the previous section we found a trivial spectrum on the GO background assuming the section condition $\tilde\p^\mu=0$, which matches the GO spectrum \req{eq:GO-spectrum} for zero winding number.
It is also natural to consider the T-dual section condition: $\tilde\p^\mu = 0$ for $\mu\ne 1$ and $\p_1 = 0$.
In the T-dual frame the winding number $w$ becomes the dual momentum $\tilde n$, and so we expect the GO spectrum $N=\tilde N=1$ to contain the non-trivial states
\begin{equation}
E = \frac{\mu\tilde n R}{\alpha'} + \frac{\alpha'k^2}{2\tilde nR} .
\label{eq:dual-GO-disp}
\end{equation}
T-duality in DFT can be performed by conjugating by elements of $\ODD$.
In the $\alpha$ sector the matrix in question has the form
\begin{equation}
(\cO_A^{\phantom AB}) = \left( \ba{cccc} 1 & 0 & 0 & 0 \\ 0 & 0 & 0 & 1 \\ 0 & 0 & 1 & 0 \\ 0 & 1 & 0 & 0 \ea \right) .
\end{equation}
Conjugating by $\cO$, we find that the dual configuration is geometric, with metric
\begin{equation}
ds^2 = -f\, dt^2 + 2\,dt\,dx + (dx^i)^2 ,
\label{eq:LC-geom}
\end{equation}
a light-cone compactification.
Here we have set $\theta = \tilde x_1$, which has natural periodicity $\theta\sim\theta+2\pi\tilde R$.


In this configuration, the metric fluctuation $h_{MN}$ takes the form
\begin{equation}
h^{\mu\nu} = -g^{\mu\lambda}g^{\nu\sigma}h_{\lambda\sigma}\,,
\qquad
h_\mu^{\phantom\mu\nu} = b_{\mu\lambda}g^{\lambda\nu} \,,
\qquad
h^\mu_{\phantom\mu\nu} = -g^{\mu\lambda}b_{\lambda\nu}\,,
\end{equation}
with $b_{\mu\nu}$ antisymmetric.

The equations of motion can be adapted from above and read
\begin{align}
\cE_\psi &= p^\mu p^\nu h_{\mu\nu} - 4p^2\psi = 0 \,,\\
\cE_{(\mu\nu)} &= p^\lambda p_{(\mu}h_{\nu)\lambda} - p^2 h_{\mu\nu} + 2p_\mu p_\nu \psi = 0 \label{eom-sym} \,,\\
\cE_{[\mu\nu]} &= p^\lambda p_{[\mu}h_{\nu]\lambda} + p^2 b_{\mu\nu} = 0 . \label{eom-anti} 
\end{align}
Under gauge variations $\xi^M = (\tilde\lambda_\mu,\lambda^\mu)$, $\delta h_{\mu\nu} = p_\mu\lambda_\nu + p_\nu\lambda_\mu$ and
$\delta b_{\mu\nu} = p_\mu\tilde\lambda_\mu - p_\nu\tilde\lambda_\mu$, so we may choose the gauge condition $h_{0\mu} = b_{0\nu} = 0$.

The $(00)$ symmetric component reads $2p_0^2\psi = 0$, and so for a propagating mode, $\psi = 0$.
The $0i$ components then imply that the remaining tensors are spatially transverse, $p^i h_{ij} = p^i b_{ij} = 0$.
The remaining equations read $p^2h_{ij} = p^2b_{ij} = 0$, giving the mass-shell condition $p^2 = 0$.
The number of degrees of freedom is the number of components of spatially transverse tensors.
For $h_{ij}$ this is $\frac{D(D-1)}{2}-(D-1)=\frac{(D-1)(D-2)}{2}$, and for $b_{ij}$, $\frac{(D-1)(D-2)}{2}-(D-1)=\frac{(D-1)(D-4)}{2}$,
giving a total of $(D-1)(D-3)$ polarizations.

The condition $p^2 = 0$ for the metric \req{eq:LC-geom} yields the dispersion relation
\be
E = \half f p_\theta + \frac{k^2}{2p_\theta} .
\label{eq:dispersion}
\ee
The radius of the $\theta$ circle is given by $\tilde R = \alpha'/R$, and so the $\theta$ momentum is quantized in units of $1/\tilde R$, $p_\theta = \frac{\tilde n R}{\alpha'}$.
If we introduce the chemical potential $\mu = f/2$ and insert these relations into the dispersion relation \req{eq:dispersion}, we obtain the Gomis-Ooguri spectrum at $N=\tilde N=1$ given in equation \req{eq:dual-GO-disp}.
%
%

\section{Conclusions \& Outlook}\label{sec:5}

In this work we analyzed the target space dynamics of DFT around a  non-Riemannian string background corresponding to the non-relativistic closed string theory of Gomis and Ooguri~\cite{Gomis:2000bd}. This non-relativistic closed string theory is, as reviewed in section \ref{ssec:NRCS}, a certain limit of a relativistic closed string compactified on a circle in the presence of an NS-NS $B$-field. As we showed in section \ref{sec:SigmaModels}, the sigma model description of \cite{Gomis:2000bd} can be embedded into the DFT sigma model of \cite{Lee:2013hma} where  the Lagrange multipliers $\beta$, $\bar\beta$ of \cite{Gomis:2000bd} are identified with components of the  vector potential of  \cite{Lee:2013hma} implementing the coordinate gauge transformations of the doubled-yet-gauged spacetime. The generalized metric \eqref{dualgeom2} corresponding to  this embedding is well-defined within DFT, but does not admit the usual decomposition in terms of metric and $B$-field of equation~\eqref{nondegH}. Thus, it provides  an example of a locally non-Riemannian  background of closed string theory.\footnote{This locally non-geometric nature is distinct from T-folds and similar backgrounds, which are locally geometric but globally not: there, string duality transformations are used to glue together locally geometric descriptions.} The fact that the DFT sigma model of \cite{Lee:2013hma} reduces to the closed string sigma model of \cite{Gomis:2000bd} hence constitutes a nontrivial check of the validity of DFT, which goes beyond the purview of Riemannian geometry. 

In section \ref{sec:4} we then analyzed the spectrum of linear perturbations around this non-Riemannian DFT background, which describes the NS-NS sector of closed string theory. This is the main result of the present  paper. We in particular showed  (in section \ref{ssec:GOspectrum}) that, in accordance with \cite{Gomis:2000bd}, there are no perturbative propagating degrees of freedom in the Kaluza-Klein sector.
We furthermore showed in section \ref{ssec:Tdualspec} that the spectrum of winding modes correctly reproduces the non-relativistic excitation spectrum found in \cite{Gomis:2000bd}. On the way, we also derived the explicit realization of the Bargmann algebra on the target space DFT manifold (section \ref{sec:galilean-symmetry}), and
presented a novel DFT geometry with Schr\"odinger conformal symmetry (section \ref{ssec:Schroedinger}). 

In order to carry out the fluctuation analysis of section \ref{sec:4}, we first derived in section \ref{sec:2} a compact form of the bosonic DFT Lagrangian expanded to second order in fluctuations around a generic on-shell background  \eqref{effectL}, expressed in terms of novel completely covariantized differential operators \eqref{comp1} and \eqref{comp2}. The compact expressions \eqref{EOMf1}, \eqref{EOMf2} and \eqref{effectL}, derived in terms of the variations of the projection operators \eqref{deltaPconstraint}, enabled us to write the fluctuation equations in a simple form. We hope that the expressions \eqref{EOMf1}, \eqref{EOMf2} and \eqref{effectL} will also be useful for the analysis of the fluctuations in different physical setups, such as 
cosmological perturbation theory, or perturbations around D-branes and other solitonic objects \cite{Berkeley:2014nza,Berman:2014jsa} within the DFT framework.

We envision that both the general fluctuation analysis of section \ref{sec:2} as well as the ability of DFT to correctly describe 
non-geometric backgrounds as seen in section \ref{sec:SigmaModels} and \ref{sec:4} will have 
further applications. For example, it may be possible to relate the non-relativistic limits of D-branes and membranes already discussed in \cite{Gomis:2000bd} to new non-geometric and non-relativistic solitonic objects in DFT. In fact, recent definitions of conserved charges in DFT \cite{Blair:2015eba,Park:2015bza}  may enable the study of such solitonic objects with Ramond-Ramond charges (D-branes) directly within the DFT setup. One 
instance of non-Riemannian geometry that would be interesting to analyze from the DFT point of view is 
the non-commutative geometry emerging on D-brane world volumes in the presence of NS-NS $B$-fields \cite{Seiberg:1999vs}. 
On a related note,
it would be interesting to search for novel boundary states 
within the DFT sigma model formulation~\cite{Lee:2013hma}. 
From the point of view of holography, it will be interesting to apply DFT to (partially) non-Riemannian situations which allow for a holographic dual description. This could for example include holographic backgrounds with non-Riemannian internal spaces, the Schr\"odinger geometries derived in section \ref{ssec:Schroedinger}, and variants thereof such as 
Lifshitz or hyperscaling violating backgrounds. 
For example, one concrete application of DFT holography would be the derivation of the asymptotic symmetry algebra of the Schr\"odinger solutions \eqref{schroedinger-geometry} by means of the Brown-Henneaux procedure. We plan to return to these and related questions in the near future~\cite{WIP}.


~\\
\textbf{Acknowledgements.}  SMK wishes to thank Yoonji Suh for helpful discussion. We thank the anonymous referee for constructive suggestions. This work was  supported in part by the World Premier International Research Center Initiative (WPI), MEXT, Japan for RM and CMT, and by the National Research Foundation of Korea through the Grants   2013R1A1A1A05005747 and 2015K1A3A1A21000302 for JHP. JHP would also like to thank Kavli IPMU for the hospitality during a seminar visit, when this project was initiated. 

\newpage

\appendix

\section{Derivation of the Fluctuation Equations and Lagrangian}\label{app:A}
%
In this appendix we will show in more detail the steps involved in deriving the linearized equations of motion \eqref{EOMf1}, \eqref{EOMf2} of the DFT Lagrangian \eqref{DFTL}, focusing on the non-trivial part of the procedure. 
We will also prove that the new operators $\Delta_A^{\phantom{A}B}$ \eqref{comp1}, \eqref{comp2}, in terms of which the linearized EOMs are expressed, are indeed covariant.

The strategy is as follows: starting from DFT Lagrangian \eqref{DFTL}, we vary with respect to the DFT covariant fields ${\cal H}_{AB}$ and $d$ to obtain the DFT EOM. We then vary these EOMs once to obtain the linearized fluctuation EOMs. The starting point of this calculation is the DFT Lagrangian \eqref{DFTL}. 
Since there are two independent fields in DFT, ${\cal H}_{AB}$ and $d$, there are two independent EOMs obtained from the variations $\delta{\cal H}_{AB}$ and ${\delta d}$. The projector $P_{AB}$ is built out of of the constant metric ${\cal J}_{AB}$ and the generalized metric ${\cal H}_{AB}$, which leads to the following relations between ${\delta H}_{AB}$, $\delta P_{AB}$, and $\delta \bar{P}_{AB}$:
\be\label{ProjectorRelations}
\ba{ll}
P_{AB} = \half({\cal J} + {\cal H})_{AB}, & \bar{P}_{AB} = \half({\cal J} - {\cal H})_{AB}\,, \\
\delta P_{AB} = \half \delta{\cal H}_{AB}, & \delta \bar{P}_{AB} = -\half \delta {\cal H}_{AB}\,,\\
\delta {\cal J}_{AB} = 0\,.
\ea
\ee
The EOM of DFT are then obtained by varying \eqref{DFTL}:
\\
\be
\ba{ll} \label{DFTEOM}
\delta {\cal L} =& -\frac{1}{4} e^{-2d} \delta d (P^{AC}P^{BD} - \bar{P}^{AC} \bar{P}^{BD})S_{ABCD}\\
& +\frac{1}{8} e^{-2d}(\delta P^{AC} P^{BD} + P^{AC} \delta P^{BD} - \delta \bar{P}^{AC} \bar{P}^{BD} - \bar{P}^{AC} \delta \bar{P}^{BD})S_{ABCD}\\
& +\frac{1}{8} e^{-2d}(P^{AC}P^{BD} - \bar{P}^{AC}\bar{P}^{BD})\delta S_{ABCD}\,.
\ea
\ee
The first line in (\ref{DFTEOM}) implies that the dilaton EOM is nothing but the Lagrangian itself, which should vanish on-shell. The second line is the EOM  for the generalized metric ${\cal H}_{AB}$, and the last line is a boundary term. We are considering on-shell backgrounds only in this work, and we will not keep track of the boundary terms in the remaining calculations. Next we vary eq.(\ref{DFTEOM}) once more and collect the quadratic terms in the variations, to obtain the linearized EOM. In doing so a subtlety arises: The generalized metric ${\cal H}_{AB}$ is a constrained object, with the constraint being
\be
{\cal H}_{AC} {\cal H}^{C}_{\phantom{C}B} = {\cal J}_{AB}\,.
\ee
This specific constraint relates the quadratic variation of the generalized metric with terms quadratic in the first variation of the generalized metric. It turns out that it is convenient to work with the projectors $P$, $\bar P$, instead of the metric itself:
\be \label{varconstraint}
\delta^2 P^{AC} = (\delta P^A_{\phantom{A}I} \bar{P}^B_{\phantom{B}J} + P^A_{\phantom{A}I} \delta \bar{P}^B_{\phantom{B}J} + \delta \bar{P}^A_{\phantom{A}I} P^B_{\phantom{B}J} + \bar{P}^A_{\phantom{A}I} \delta P^B_{\phantom{B}J})\delta P^{IJ}\,.
\ee
Ignoring the boundary term in eq. (\ref{DFTEOM}) and performing another variation will then lead to the following result:
\be
\ba{ll} \label{Qvari}
\half \delta^2 {\cal L} &= -\frac{1}{8} e^{-2d} \delta d (P^{AC}P^{BD} - \bar{P}^{AC} \bar{P}^{BD}) \delta S_{ABCD}\\
&\quad +\frac{1}{16} e^{-2d}(\delta P^{AC} P^{BD} + P^{AC} \delta P^{BD} - \delta \bar{P}^{AC} \bar{P}^{BD} - \bar{P}^{AC} \delta \bar{P}^{BD}) \delta S_{ABCD} \\
&\quad + \frac{1}{16} e^{-2d}(\delta^2 P^{AC} P^{BD} + P^{AC} \delta^2 P^{BD} - \delta^2 \bar{P}^{AC} \bar{P}^{BD} - \bar{P}^{AC} \delta^2 \bar{P}^{BD})S_{ABCD}\\
&= -\frac{1}{8} e^{-2d} \delta d (P^{AC}P^{BD} - \bar{P}^{AC} \bar{P}^{BD}) \delta S_{ABCD}\\
&\quad +\frac{1}{16} e^{-2d}(\delta P^{AC} P^{BD} + P^{AC} \delta P^{BD} - \delta \bar{P}^{AC} \bar{P}^{BD} - \bar{P}^{AC} \delta \bar{P}^{BD}) \delta S_{ABCD} \\
&\quad + \frac{1}{4} e^{-2d}(\bar{P}^{AC} \bar{P}^{BD} - P^{AC} P^{BD})S_{AB} \delta P_{CE} \delta P^{E}_{\phantom{E}D}\,.
\ea
\ee
Note that arriving at the third line of the eq. (\ref{Qvari}) required the use of eq. (\ref{varconstraint}). The remaining computation is tedious but straightforward, resulting in
\be
\ba{ll} \label{FQvari}
\half \delta^2 {\cal L} =& \frac{1}{16} e^{-2d}[\{\bar{P}_{EF}(\bar{P}^{AB} P_{CD} - P^{B}_{\phantom{B}C}P^{A}_{\phantom{A}D}) + P_{EF}(\bar{P}^{B}_{\phantom{B}C} \bar{P}^{A}_{\phantom{A}D} - P^{AB} \bar{P}_{CD})\}\nabla_A \delta P^{CE} \nabla_B \delta P^{DF}\\
& + 8(P^{AE} -\bar{P}^{AE})\partial_A \delta d \partial_E \delta d - 8 \partial_A \delta d \nabla^E \delta P^{A}_{\phantom{A}E} + 4(\bar{P}^{AC} \bar{P}^{BD} - P^{AC} P^{BD})S_{AB} \delta P_{CE} \delta P^{E}_{\phantom{E}D}]\,.
\ea
\ee
In terms of  our new  second order differential  operators,  $\Delta_A^{\phantom{A}B},\brDelta_A^{\phantom{A}B}   $~(\ref{DeltabrDelta}), the above expression  can be simplified dramatically,
\be
\half \delta^2 {\cal L} = e^{-2d}\left[\half(P-\brP)^{AB}\partial_{A}\delta d\,\partial_{B}\delta d-\half\partial_{A}\delta d\,\na_{B}\delta P^{AB}+\textstyle{\frac{1}{8}}\delta P^{AB}(\brDelta_{A}{}^{C}P_{B}{}^{D}-\Delta_{A}{}^{C}\brP_{B}{}^{D})\delta P_{CD}\right].
\ee
This result coincides exactly with (\ref{effectL}).

The final result of the linearized EOM should be covariant under the generalized DFT diffeomorphisms. This implies that the difference between generalized diffeomorphisms $\delta_X$ and the corresponding generalized Lie derivative ${\cal L}_X$ must vanish. The first and second terms of eq. (\ref{FQvari}) are rather trivially seen to be covariant by imposing the relation (\ref{anomalous}). However the last part including the $\bar{\Delta}_{A}^{\phantom{A}B}$ operator is not obviously covariant, since $\bar{\Delta}_{A}^{\phantom{A}B}$ contains second order semi-covariant derivatives. We  will now show that the following second-order derivative is  completely covariant, 
\be \label{tterm}
P_{I_1}^{\phantom{I_1}A_1}
\cdots P_{I_{i-1}}^{\phantom{I}A_{i-1}}
\bar{\Delta}_{I_i}^{\phantom{I_i}G} P_{I_{i+1}}^{\phantom{I}A_{i+1}}\cdots P_{I_n}^{\phantom{I_n}A_n} T_{A_1 \cdots A_{i-1}GA_{i+1} \cdots A_n} \,.
\ee
Here the subscript index $i$ merely indicates the  position of the contracted $\ODD$ index, $G$, and can be taken arbitrary among  $1,2,\cdots, n$.  To compute the potential violation of covariance, $(\delta_X - {\cal L}_X)$, we  recall the generic expression of the anomalous terms for the second order semi-covariant derivative,
\be
\ba{ll}
\dis{(\delta_{X}{-\hcL_{X}})\na_{B}\na_{C}T_{A_{1}\cdots A_{n}}}=&
2(\cP{+\brcP})_{BC}{}^{GFDE}
\partial_{F}\partial_{[D}X_{E]}\na_{G}T_{A_{1}\cdots A_{n}}\\
{}&+
\sum_{i=1}^{n}2(\cP{+\brcP})_{BA_{i}}{}^{GFDE}
\partial_{F}\partial_{[D}X_{E]}\na_{C}T_{A_{1}\cdots A_{i-1} GA_{i+1}\cdots A_{n}}\\
{}&+
\sum_{i=1}^{n}2(\cP{+\brcP})_{CA_{i}}{}^{GFDE}
\partial_{F}\partial_{[D}X_{E]}\na_{B}T_{A_{1}\cdots A_{i-1} GA_{i+1}\cdots A_{n}}\\
{}&+
\sum_{i=1}^{n}2(\cP{+\brcP})_{CA_{i}}{}^{GFDE}
\left(\na_{B}\partial_{F}\partial_{[D}X_{E]}\right)T_{A_{1}\cdots A_{i-1} GA_{i+1}\cdots A_{n}}\,.
\ea
\label{noncov2}
\ee
From this expression we obtain straightforwardly
\be
\ba{l} \label{covRes}
(\delta_X - {\hcL}_X)P_{I_1}^{\phantom{I_1}A_1}
\cdots P_{I_{i-1}}^{\phantom{I}A_{i-1}}
\bar{\Delta}_{I_i}^{\phantom{I_i}G} P_{I_{i+1}}^{\phantom{I}A_{i+1}}\cdots P_{I_n}^{\phantom{I_n}A_n} T_{A_1 \cdots A_{i-1}GA_{i+1} \cdots A_n} \\ 
= 4 \bar{\cal P}^{B\phantom{I_i}GDEF}_{\phantom{B}I_i}\partial_D \partial_{[E} X_{F]} \nabla_B T_{A_1 \cdots A_{i-1}GA_{i+1} \cdots A_n}P_{I_1}^{\phantom{I_1}A_1}\cdots P_{I_n}^{\phantom{I_n}A_n}\\
\quad +2 \bar{\cal P}^{B\phantom{I_i}GDEF}_{\phantom{B}I_i} \nabla_B  \partial_D \partial_{[E} X_{F]} T_{A_1 \cdots A_{i-1}GA_{i+1} \cdots A_n}P_{I_1}^{\phantom{I_1}A_1}\cdots P_{I_n}^{\phantom{I_n}A_n}\\
\quad -4 \bar{\cal P}^{G\phantom{I_i}BDEF}_{\phantom{G}I_i}\partial_D \partial_{[E} X_{F]} \nabla_B T_{A_1 \cdots A_{i-1}GA_{i+1} \cdots A_n}P_{I_1}^{\phantom{I_1}A_1}\cdots P_{I_n}^{\phantom{I_n}A_n}\\
\quad -4 \bar{\cal P}^{\phantom{I_i}BGDEF}_{I_i}\nabla_B \partial_D \partial_{[E} X_{F]} T_{A_1 \cdots A_{i-1}GA_{i+1} \cdots A_n}P_{I_1}^{\phantom{I_1}A_1}\cdots P_{I_n}^{\phantom{I_n}A_n}\\
\quad -4 \bar{\cal P}^{\phantom{I_i}BGDEF}_{I_i}\partial_D \partial_{[E} X_{F]} \nabla_B T_{A_1 \cdots A_{i-1}GA_{i+1} \cdots A_n}P_{I_1}^{\phantom{I_1}A_1}\cdots P_{I_n}^{\phantom{I_n}A_n}\\
\quad +2 \bar{\cal P}^{GB\phantom{I_i}DEF}_{\phantom{GB}I_i}\nabla_B \partial_D \partial_{[E} X_{F]} T_{A_1 \cdots A_{i-1}GA_{i+1} \cdots A_n}P_{I_1}^{\phantom{I_1}A_1}\cdots P_{I_n}^{\phantom{I_n}A_n}\\
\quad +2 \bar{\cal P}^{\phantom{I_i}GBDEF}_{I_i}\nabla_B \partial_D \partial_{[E} X_{F]} T_{A_1 \cdots A_{i-1}GA_{i+1} \cdots A_n}P_{I_1}^{\phantom{I_1}A_1}\cdots P_{I_n}^{\phantom{I_n}A_n}\,,
\ea 
\ee
which further simplifies to show the desired covariant property,  
\be
\ba{l} \label{FcovRes}
(\delta_X - {\hcL}_X)P_{I_1}^{\phantom{I_1}A_1}
\cdots P_{I_{i-1}}^{\phantom{I}A_{i-1}}
\bar{\Delta}_{I_i}^{\phantom{I_i}G} P_{I_{i+1}}^{\phantom{I}A_{i+1}}\cdots P_{I_n}^{\phantom{I_n}A_n} T_{A_1 \cdots A_{i-1}GA_{i+1} \cdots A_n}  \\
= 6\cJ_{I_{i}J}\bar{{\cal P}}^{[JBG]DEF}\nabla_B \partial_D \partial_{[E} X_{F]} T_{A_1 \cdots A_{i-1}GA_{i+1} \cdots A_n}P_{I_1}^{\phantom{I_1}A_1}\cdots P_{I_n}^{\phantom{I_n}A_n}\\
= 6\cJ_{I_{i}J}\bar{{\cal P}}^{[JBG][DEF]}\nabla_B \partial_{[D} \partial_{E} X_{F]} T_{A_1 \cdots A_{i-1}GA_{i+1} \cdots A_n}P_{I_1}^{\phantom{I_1}A_1}\cdots P_{I_n}^{\phantom{I_n}A_n}\\
=0\,.
\ea
\ee
Similarly, one can show that the other operator, $\brP_{I_1}^{\phantom{I_1}A_1}\cdots {\Delta}_{I_i}^{\phantom{I_i}G}\cdots \brP_{I_n}^{\phantom{I_n}A_n} T_{A_1 \cdots A_{i-1}GA_{i+1} \cdots A_n}$, is completely covariant as well.



\end{document}